\newcommand\T{\rule{0pt}{2.6ex}}       
\newcommand\B{\rule[-1.2ex]{0pt}{0pt}} 
\documentclass{aa}  

\usepackage{graphicx}
\usepackage{natbib}

\bibpunct{(}{)}{;}{a}{}{,} 
\usepackage{txfonts}
\begin{document}

   \title{Revealing the large nuclear dust structures in NGC 1068 with MIDI/VLTI  \thanks{Based on observations made with ESO Telescopes at the La Silla Paranal Observatory under programme ID 080.B-0928 and 089.B-0099.  Based on data obtained from the ESO Science Archive Facility.}}

   \author{N. L\'opez-Gonzaga\inst{1}\and W. Jaffe\inst{1}
          \and L. Burtscher\inst{2} \and K. R. W. Tristram\inst{3} \and K. Meisenheimer\inst{4}
          }

   \institute{Sterrewacht Leiden, Universiteit Leiden, Niels-Bohr-Weg 2, 2300 CA Leiden, The Netherlands  \\ \email{nlopez@strw.leidenuniv.nl}
         \and 
         Max-Planck-Institut f\"{u}r extraterrestrische Physik, Postfach 1312, Gie\ss enbachstr., 85741 Garching, Germany
         \and
         Max-Planck-Institut f\"{u}r Radioastronomie, Auf dem H\"{u}gel 69, 53121 Bonn, Germany
         \and 
         Max-Planck-Institut f\"{u}r Astronomie, K\"{o}nigstuhl 17, 69117 Heidelberg, Germany
             }

   \date{Received ;}

  
  \abstract
   {}
   { To understand the relation in AGNs between the small 
   ``obscuring torus" and dusty structures at larger scales (5-10 pc).} 
   {The dusty structures in AGNs are best observed in the mid-infrared.  To
   achieve the necessary spatial resolution ($\sim 20-100$ millarcsec)
   we use ESO's Mid-Infrared Interferometer (MIDI) with the 1.8 m Auxiliary Telescopes.  We use the {\it chromatic phases} in the data to improve the spatial fidelity of the analysis.}
   {We present interferometric data for NGC 1068 obtained in 2007 and 2012. We find no evidence of source variability. Many
   $(u,v)$ points show non-zero chromatic phases indicating significant asymmetries. 
   Gaussian model fitting of the correlated fluxes and chromatic phases 
   provides a 3-component best fit with estimates of sizes, temperatures and positions of the components. 
   A large, warm, off-center component is required at a distance approximately 90 mas to the north-west at a PA $\sim -18^\circ$.  }
   {The dust at 5-10 pc in the polar region contributes 4 times more to the mid-infrared flux at 12 $\mu$m than the dust located at the center. This dust may represent the
   inner wall of a dusty cone. If similar regions are heated by the direct 
   radiation from the nucleus, then they will contribute substantially to
   the classification of many Seyfert galaxies as Type 2. Such a region is also consistent in other Seyfert galaxies (the Circinus galaxy, NGC 3783 and NGC 424).  }

   \keywords{techniques: interferometric -- galaxies: active -- galaxies: nuclei -- galaxies: Seyfert -- galaxies: individual: NGC 1068 -- radiation mechanisms: thermal}
   
   \maketitle
%

\section{Introduction.}

Active Galactic Nuclei (AGN) have been intensely studied because they host many interesting physical process, such as accretion of material and formation of jets. Many subclasses of AGNs have been defined based on observational criteria; the earliest of these was defined by Seyfert \citep{1943ApJ....97...28S} by the presence of high ionization forbidden lines. They additionally show low ionization lines and very high ionization coronal lines. The similar line ratios from galaxy to galaxy suggest that they are powered by the same type engine. But they also show differences that have led to a dual classification: Type 1 galaxies show broad optical permitted lines absent in Seyfert Type 2. The idea that Type 1 and Type 2 share the underlying engine is strongly supported by the detection of polarized broad lines in  3C 234 \citep{1984ApJ...278..499A} and NGC 1068 \citep{1985ApJ...297..621A}. These polarized spectra, most probably caused by electron scattering, revealed the existence of a Broad Line Region (BLR) in Type 2 galaxies. 

To explain the different appearances of various types of AGN, the existence of an axisymmetric dusty structure, a {\it torus}, was proposed in the context of AGN Unified Models \citep{1993ARA&A..31..473A, 1995PASP..107..803U}.  The general concept is that the Type 2  galaxies in fact are absorbed Type 1 galaxies, where the orientation and absorption of the torus play a mayor role shaping the apparent properties. 
The  energy absorbed by the torus will be re-emitted mainly in the mid-infrared wavelength regime, giving rise to a pronounced peak in the spectral energy distribution of many AGN \citep{1989ApJ...347...29S}. Resolving the morphology of this mid-infrared radiation is the key to understanding the physical properties of the dust structures.  However they are typically too small to be resolved even with the largest single dish telescopes. Only with the availability of powerful techniques such as mid-infrared interferometry further progress has been possible.  Several interferometric studies in the mid-infrared have been published for individual galaxies. They include the brightest AGNs: the Circinus galaxy \citep{2007A&A...474..837T}, NGC 1068 \citep{2004Natur.429...47J,2006A&A...450..483P, 2009MNRAS.394.1325R}, Centaurus A \citep{2007A&A...471..453M,2010PASA...27..490B} and the brightest Type 1 Seyfert galaxy, NGC 4151 \citep{2009ApJ...705L..53B}. Recently two fainter sources,  NGC 424 
\citep{2012ApJ...755..149H} and NGC 3783 
\citep{2008A&A...486L..17B, 2013ApJ...771...87H} were observed with a very well sampled $(u,v)$ coverage. Studies with the intention of getting general properties of the objects are also published. \citet{2009A&A...493L..57K} claimed evidence for a 'common radial structure'  for the nearby AGN tori, \citet{2007A&A...474..837T} demonstrated that weak AGNs can also be observed with MIDI and saw first evidence for a size-luminosity relation \citep{2009A&A...502...67T}. \citet{2013A&A...558A.149B} modelled 23 AGNs and found that there is a large diversity in nuclear mid-IR structures that is not attributable to luminosity of the source or resolution of the observations.

This paper extends our previous work on the near-nuclear, parsec scale dust structures in
NGC~1068 in order to investigate the connection with the larger scale structures. We achive this by making use of low spatial frequency interferometric observations. The outline of this paper is the following: In Sect.~\ref{sec:background} we give a summary of the previous  mid-infrared observations of the nuclear dust in NGC~1068. We describe the new observations and data reduction process in Sect.~\ref{sec:currentstatus}. In Sect.~\ref{sec:results} we present the interferometric data with emphasis on the {\it chromatic phases} which give insights into asymmetric morphologies. We investigate the radial profile of the correlated fluxes and the possibility for variability. In Sect.~\ref{sec:gaussfits} we explain the gaussian model used to reproduce the interferometric data and the parameters that best fit. We discuss the best models in Sect.~\ref{sec:discussion}, analyse the properties of the components of the model and  identify the dust regions associated with each component.  We study in Sect.~\ref{sec:energetics} the possible heating mechanism for the two mid-infrared northern components found 
from the modelling. In Sect.~\ref{sec:asymmetry} we discuss the asymmetry of the mid-infrared nuclear region in NGC~1068 and its implications. Finally, we present our conclusions in Sect.~\ref{sec:conclusions}.

\section{Previous infrared Observations of the nucleus of NGC~1068}
\label{sec:background}

NGC 1068, at a distance of only 14.4 Mpc,  is a prototype Seyfert 2 galaxy that has been intensively studied. Its proximity and infrared brightness make it a suitable target to study the dusty structures that obscure the nucleus. Previous high spatial resolution single telescope studies revealed the existence of an infrared extended emission region around the central engine (\citet{1998ApJ...504L...5B,2000AJ....120.2904B,2001ApJ...557..637T,2005MNRAS.363L...1G} in the MIR, and \citet{1998A&A...339..687R,2004A&A...417L...1R,2006A&A...446..813G} in the NIR). In the  mid-infrared regime, single dish observations indicate that  the extended emission  has an elongation of about 1'' in the north-south direction and is unresolved in the east-west direction \citep{2000AJ....120.2904B}. The emission shows a strong asymmetry, with a larger emission area extending more to the north than to the south. 

\citet{2004Natur.429...47J} demonstrated the existence of a central parsec-sized circumnuclear dust structure in NGC 1068 using mid-infrared ($\lambda=8-13 \mu$m) interferometric observations from ESO's VLTI/MIDI. \citet{2009MNRAS.394.1325R} reported further MIDI observations with a more extensive $(u,v)$ coverage of sixteen baselines which allowed them to investigate the structure of the inner regions of the obscuring disk with greater detail. In both cases, a two component model, each with a gaussian brightness distribution,  was used to fit the correlated fluxes obtained from MIDI. The size and orientation of the hot component ($\sim 800$~K), associated to the inner funnel of the obscuring disk, were well fitted with an elongated gaussian 1.35 parsec long  and 0.45 parsec thick (FWHM)  at PA= -42$^\circ$. The data strongly suggest that the dusty disk and the optical ionization cones from the jet are misaligned in NGC 1068. The disk is in fact co-linear with the H$_2$O megamaser disk \citep{1996ApJ...472L..21G}.

A second, more extended, component was also detected. This component was over-resolved by the interferometer and its geometrical parameters were not well constrained. The analysis implied a  warm ($\sim 300$~K) structure of $\sim 3 \times 4$ pc size. While the position angle could not be determined, the authors suggest that a north-south elongated structure could be identified with the elongation of the mid-infrared region of NGC 1068, seen by \citet{2000AJ....120.2904B}, who attribute it to re-emission by dust of UV radiation concentrated in the ionization cone. This component is part of the environment surrounding the inner hot dust region, and according to \citet{2007A&A...472..823P}, it represents a large fraction of the emission within the MIDI field of view. Using single telescope VISIR data, they find a compact component $<$ 85 milliarcseconds (mas) in size directly associated with the dusty torus, and an elliptical component of size ($<$ 140) mas $\times$ 1187 mas at PA $\sim -4^\circ$. They suggest  
that the extended environment surrounding the compact 800 K dust region contributes more than 83 \% of the total core emission.

\section{The current observations}
\label{sec:currentstatus}

\subsection{Motivation}

Since the extended component was overresolved  in the observations reported  by \citet{2009MNRAS.394.1325R}, little is known about the physical nature of the structures on 5 - 10 parsec scales and  different models could describe this region. The cooler emission on these scales may simply represent an extension of the inner dust accretion disk at larger scales. It may also arise in the  intermediate region
between the inner dust accretion disk and the outer circumnuclear starforming regions as suggested by the co-evolution scenario of nuclear starbursts and tori from \citet{2008A&A...491..441V} and modelled for the case of NGC 1068 by \citet{2009MNRAS.393..759S,2010MNRAS.403.1801S}. We may also have a region where interactions between the accreting dust structures and infalling material \citep{2009ApJ...691..749M} and winds originating near the nucleus are present. To clarify these questions we obtained a new set of mid-infrared interferometric observations with MIDI/VLTI, using smaller baselines to
better map these larger scale components.
   
   \begin{figure}
   \centering
   \includegraphics[width=\hsize]{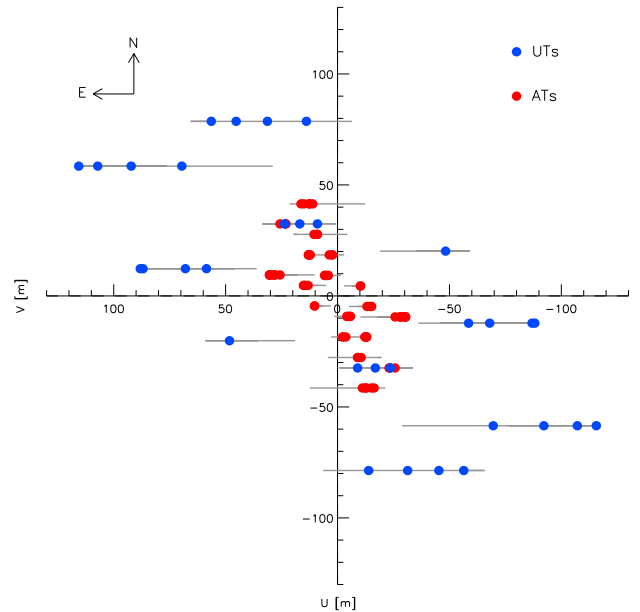}
      \caption{$(u,v)$ coverage for NGC 1068 showing both UTs and ATs configurations.
Blue dots show the measurements taken with the UTs and red dots represent
the $(u,v)$ points measured with the ATs in 2007 and 2012.     
              }
         \label{Figuvflx}
   \end{figure}
%

  \subsection{Description of the observations} 
  
Our interferometric observations were performed in the $N$ band in a wavelength
range from 8 to 13 $\mu$m with the MID-infrared Interferometric instrument 
\citep[MIDI,][]{2003Ap&SS.286...73L} at the Very Large Telescope Interferometer
(VLTI) located on Cerro Paranal in Chile and operated by the European Southern
Observatory (ESO).   

MIDI is a two beam Michelson interferometer that combines the light from two
8.2-meter Unit Telescopes (UTs) or two 1.8-meter Auxiliary Telescopes (ATs). The
main observables from MIDI are the single-dish spectra and the correlated flux
spectra that are obtained from the interference pattern generated by the two beams. 
For our new observations we used only the ATs. They are movable, allowing the
observation of more and shorter baselines than can be observed with the UTs. Their
adequate sensitivity and available baselines from 10 to 50 meters
makes them suitable to study the region of $1 - 10$ parsec of NGC 1068. For our
observations we used the low resolution NaCl prism with spectral resolution $R\equiv
\lambda/\Delta \lambda \sim 30$ to disperse the light of the beams.
   
Observations were carried out in the nights of October 7 and 8, 2007 and  September
19, 20 and 23-26, 2012 using guaranteed time observations (GTO). Two nights of
observation (September 23 and 26) were discarded due to bad weather conditions.
A log of the observations and instrument parameters can be found in 
Appendix B. Due to the near-zero 
declination of NGC 1068 the baseline tracks in the $(u,v)$-plane are parallel
to the u-axis (c.f. Fig.~\ref{Figuvflx}). This figure shows the previous UT
observations together with the new AT observations. During our observation period we
also tried the new MIDI + PRIMA FSU mode
\citep{2010SPIE.7734E..60M,2012SPIE.8445E..0QP}. This was done for two main reasons:
1) to stabilize the fringes on the long baselines and 2) to use the FSU to get an
estimate of the K-band visibility on new $(u,v)$ points. Previous VLTI/VINCI
observations of \citet{2004A&A...418L..39W} are available for one baseline along PA
= 45$^\circ$. Unfortunately the PRIMA FSU was not sensitive enough to improve upon the
self-fringe tracking by MIDI. 

  \subsection{Calibration and Data reduction}
  
The calibrators used were HD10380 and HD18322. We choose them to be close in airmass
to the target with $\Delta (\sec z) \leq 0.25$. We started each observation night
using HD10380 as calibrator  and when the altitude of the calibrator was less than
the altitude of NGC 1068, we changed to HD18322, which at that point was located  
$~10\,^{\circ}$ higher than NGC 1068. 

We have applied  the techniques developed  during the MIDI AGN Large Program
\citep{2012SPIE.8445E..1GB} to plan our observing strategy, data reduction process
and analysis of the data. Based on their experience we have optimized our observing
sequence by switching as fast as possible between target and calibrator fringe
track. This was done by omitting single-dish observations and also avoiding fringe
searches.  For each $(u,v)$ point we performed a sequence of CAL-SCI-CAL, i.e.,
calibration measurements were taken just before and after a science fringe track,
this allowed us to have a much better estimate for the correlated flux of NGC 1068
than using standard observing procedures (CAL-SCI). The additional
calibration observations allow more reliable estimates of the instrumental
visibility and therefore of the calibrated correlated flux. To correct for correlation losses due to
atmospheric phase jitter we performed {\it dilution} experiments similar to those
done for the MIDI AGN Large Program \citep{2013A&A...558A.149B}. Correlation losses for 
our faintest fluxes are less than 10\% of the correlated fluxes, which is less than the uncertainties (see Sect.~\ref{subsec:corrfluxes})
The reduction of the data was performed with the interferometric data reduction
software MIDI Interactive Analysis and Expert Work Station 
\citep[MIA+EWS\footnote{EWS is
available for download from:\hfil\break http://home.strw.leidenuniv.nl/$\sim$jaffe/ews/index.html.},][]{2004SPIE.5491..715J}  which implements the method of coherent
integration for MIDI data. Calibration of the correlated fluxes was computed by
dividing the correlated fluxes of the target by those of the calibrator and
multiplying by the known flux of the calibrator. For HD10380 and HD18322 we used the
spectral template of \citet{1999AJ....117.1864C}. In the remainder of this paper we follow the radio astronomical custom of using {\bf correlated} fluxes
rather than {\bf visibilities} which are defined as the correlated flux divided by
the total or {\bf photometric} flux.  At short infrared wavelengths visibilities are
less susceptible to changes in atmospheric conditions than correlated fluxes, but at
longer wavelengths, i.e. in the mid-infrared, the difficulties of measuring
photometric fluxes against the fluctuations of the bright sky favor the use of
correlated fluxes.
 
   \begin{figure*}
   \centering
   \includegraphics[angle=90, width=\textwidth]{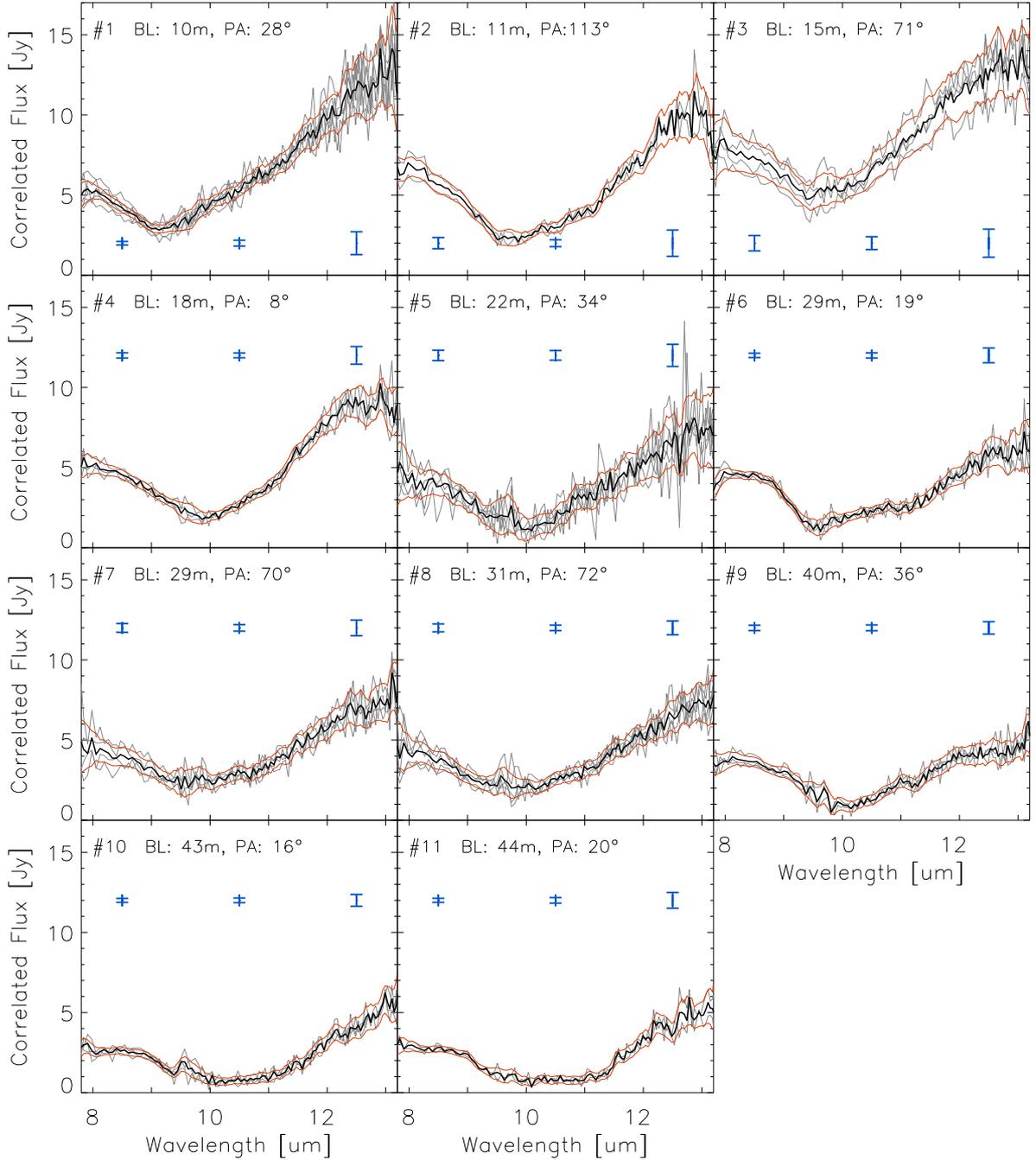}
      \caption{Amplitudes of correlated fluxes measured with the ATs and grouped by
their separation in the $(u,v)$ plane (see text for the selection criterion).  The group
numbers are given in the top left corner.  The
different correlated fluxes are display in grey lines and the average computed
spectrum is shown with a black line. The  red lines represent the region of
the 1-sigma uncertainty of a single observation. Blue bars represent the
2-sigma uncertainty of the average computed spectrum at 8.5, 10.5 and 12.5
$\mu$m. The region between $9$ and $10$ $\mu$m has higher uncertainty because
of the atmospheric O$_{3}$ absorption feature in this region.}
         \label{Figcorr}
   \end{figure*}
%

\section{Results}
\label{sec:results}

   \begin{figure*}
   \centering
   \includegraphics[angle=90, width=\textwidth]{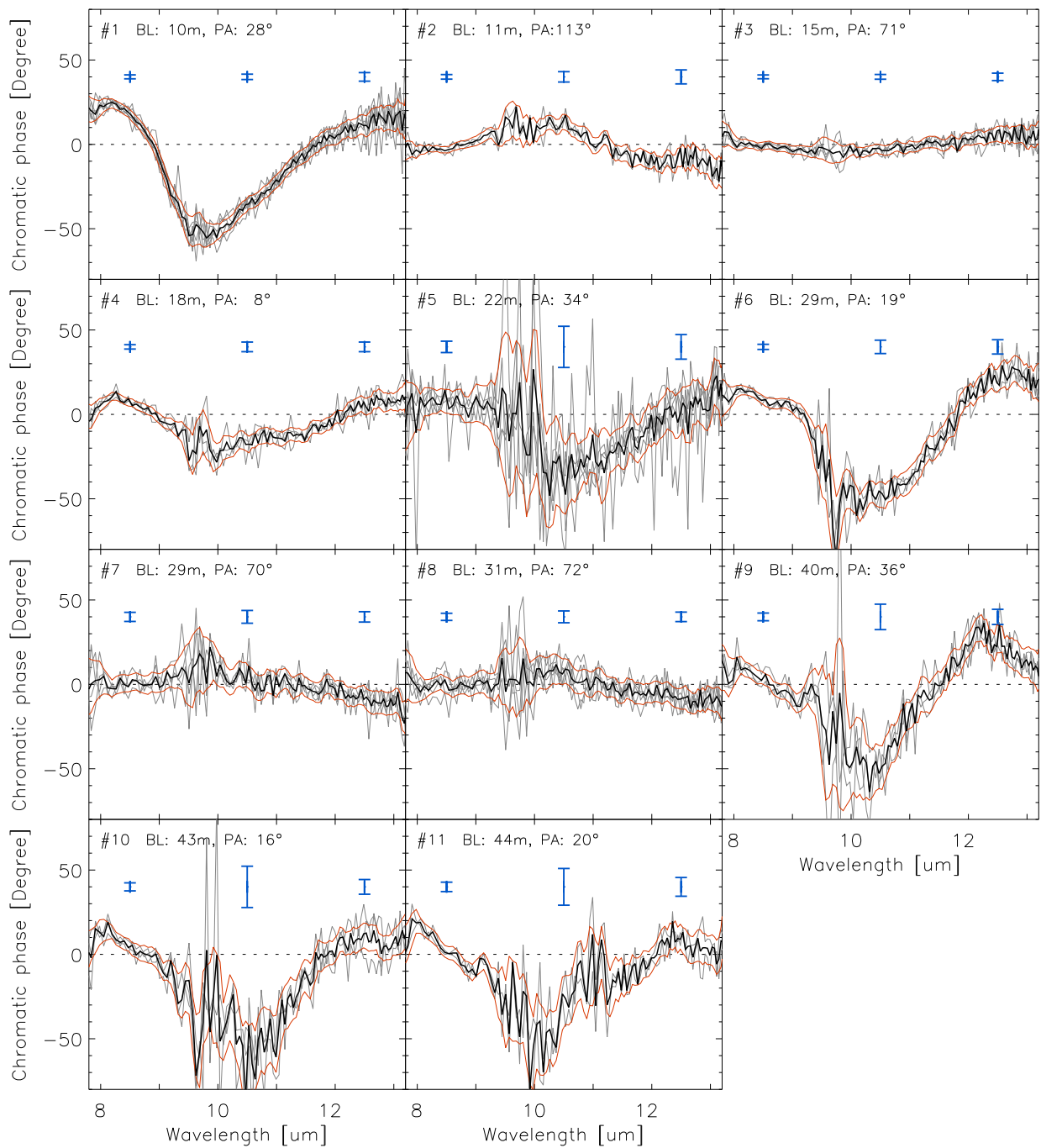}
      \caption{Plots of chromatic phases grouped by their separation in the $(u,v)$
plane.  The chromatic phases of each independent observation in the group
are given in grey lines and the average computed signal is shown with a black line. The 
red lines represent the region of the 1-sigma uncertainty of a single
observation.  Blue bars represent the 2-sigma uncertainty of the average
chromatic phase at 8.5, 10.5 and 12.5 $\mu$m.}
         \label{Figphi}
   \end{figure*}
%

\begin{figure*}
\centering
\begin{minipage}{.5\textwidth}
  \centering
  \includegraphics[angle=90,width=\linewidth]{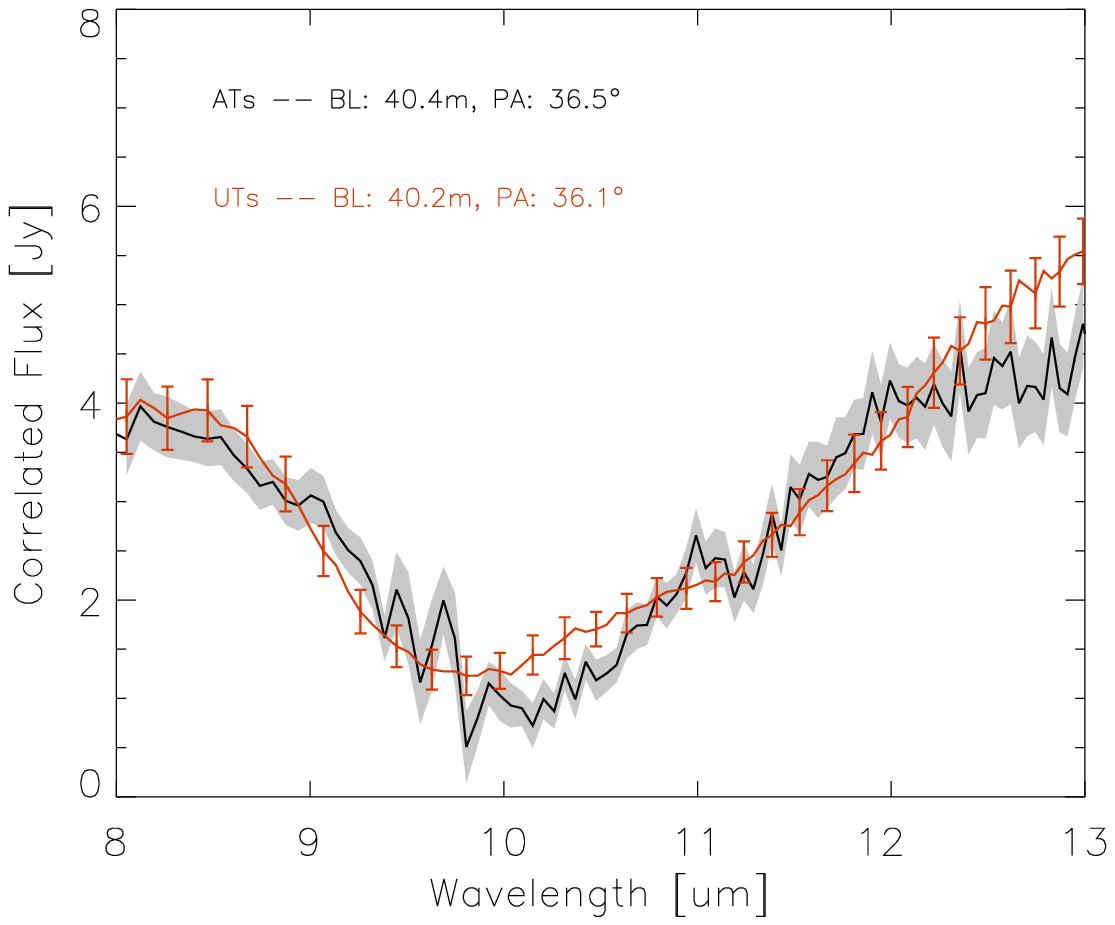}
\end{minipage}%
\begin{minipage}{.5\textwidth}
  \centering
  \includegraphics[angle=90,width=\linewidth]{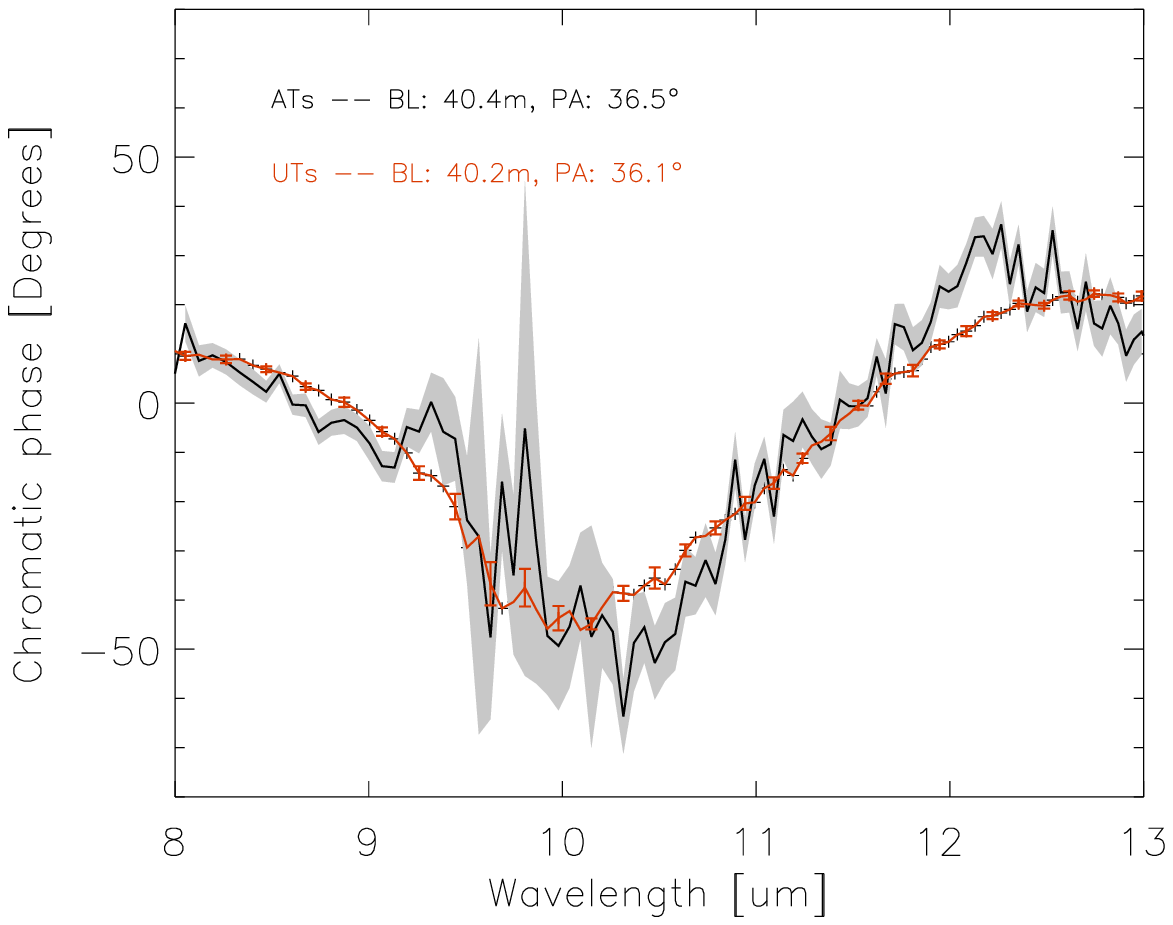}
\end{minipage}
\caption{Comparison of the correlated flux observed with a projected baseline of $
40$ m and PA=36$^{\circ}$ using UTs (observed in 2005, \citet{2009MNRAS.394.1325R})
and using ATs (observed in 2012). The red line with error bars represents the
correlated flux obtained with the UTs and black line with a grey shade region
represents the correlated flux obtained with ATs.
              }
\label{Figcheck}
\end{figure*}

\subsection{Correlated fluxes}
\label{subsec:corrfluxes}

In total, 40 correlated fluxes measured under good weather conditions were reduced
and calibrated\footnote{The same stacking method was applied to the fringe tracks as in \citet{2013A&A...558A.149B}. Fringe tracks were reduced together when they were less than 30 min apart and were calibrated with the same star.}. We have divided those visibility points into 11 groups using the
criterion that  visibility points of the same group be located within the AT diameter,
1.8 meters, of each other.  Figure~\ref{Figcorr} shows the correlated fluxes for each group,
sorted by baseline length. The
group number is indicated in the top left corner of each plot as a reference for the
discussion below. The plots include (1) spectra of the individual measurements
(grey); (2) the average of the measurements in the group (black); (3) the mean formal errors (average of the individual formal errors from EWS) (red) and (4) the formal
errors in the means (blue error bars).

To check the consistency of calibrated  interferometric fluxes with
different baselines, telescopes, under different atmospheric conditions and in
different epochs, we have taken multiple, independently calibrated measurements of the target at equivalent $(u,v)$ positions. Fluxes measured
at two adjacent $(u,v)$ points cannot differ significantly if $L\Delta u/\lambda \ll
1$, where $L$ is the overall source angular size, and $\Delta u$ the separation in
the $(u,v)$ plane. A single telescope of diameter $D$ is only sensitive to emission
with in   $L \lesssim \lambda / D$, so we conclude that two points are equivalent if
$\Delta u < D$. In our case $D=1.8$ m. If the source is in fact smaller, $L\ll
\lambda /D$, then $(u,v)$ points separated by larger than $D$ should still yield the
same flux.

For the spectra shown in Fig.~\ref{Figcorr}, we observe that all correlated fluxes
fall inside the 1-sigma uncertainty or very close to that, thus verifying the formal
estimates. The flux uncertainties in a single independent measurement at 8.5, 10.5 and
12.5 $\mu$m are typically of the order of $13\%$, $20\%$ and $17\%$; uncertainties
vary depending on the weather conditions. Even when observations of equivalent {\it
$(u,v)$} points where taken in different days, under different weather conditions,  the
correlated fluxes are consistent  with each other. Computing the average of the
measurements (see Sect.~\ref{subsec:corrfluxes}) should give us a proper estimate of the correlated flux and we can
lower the uncertainty of the error by a factor of $\sqrt{N}$, where $N$ is the
number of visibility points used to compute the average. The uncertainties for the
average computed flux are of the order of $6\%$, $11\%$ and $8\%$  at 8.5, 10.5 and 12.5 $\mu$m respectively. 

\subsection{Chromatic phases}
\label{subsec:chromaticphases}

EWS gives not only the amplitude of the (complex) source visibilities but also the
\textit{chromatic phases}. The chromatic phases are identical to the \textit{true}
interferometric phases except that the constant and linear dependencies of phase on
wavenumber  $k\equiv 2\pi/\lambda$ have been removed. This occurs because the fluctuations
in the atmospheric refractivity introduce phase shifts that are linear functions of
wavenumber. In the absence of a phase-stable external fringe tracker the removal of
these atmospheric fluctuations in the reduction process inevitably removes also the
linear components of the true source phases. This leaves only the
2\textsuperscript{nd} and higher order phase components. Chromatic phases cannot be
used directly in image reconstruction, but still constrain the source structure.
Most directly, inversion symmetric sources, averaged over the entire wavelength band, will always have zero chromatic phase.
\footnote{This description does not include cases where we have phase jumps of 180$^\circ$ at the nulls of the visibility produced by distributions such as uniform disks or rings.}

Thus non-zero chromatic phases imply asymmetric structures
\citep{2007A&A...467.1093D}.

The grouped chromatic phases of MIDI measurements are given in Fig.~\ref{Figphi}. In
each group the chromatic phases of every independent observation fall almost
entirely within the 1-sigma region meaning that the observations are consistent with each other. As in Sect.~\ref{subsec:corrfluxes} a similar computation of the average chromatic phase of each group was calculated in order to obtain a proper estimate. The typical uncertainties for the measured 
chromatic phases of an independent observation  at 8.5, 10.5 and 12.5 are 4$^\circ$ ,
10$^\circ$ and 7$^\circ$ respectively. 

The $(u,v)$ points labelled \#2, \#3, \#7, \#8, which are located at position angles between
69$^\circ$ and 114$^\circ$, show no chromatic phases, while points which are located
between 7$^\circ$ and 39$^\circ$ show  significant chromatic phases\footnote{From the 23 sources analyzed in  \citet{2013A&A...558A.149B} only NGC 1068 and Circinus show clearly visible non-zero chromatic phases. Circinus chromatic phases are analyzed in \cite{2013arXiv1312.4534T}}.
This is a clear evidence for non-symmetric structure. In a first approach this
suggests that the asymmetry axis is located close to the north-south
direction. We observe that within the range where the chromatic phases are observed (7$^\circ$ and 39$^\circ$) the largest amplitude of the chromatic phases is reached in the lowest projected baseline length; there is  a decrease in the amplitude around BL$\approx$20 m and then the chromatic phases increase slightly in amplitude until a BL$\approx$40 m. This change in the amplitude indicates that the  asymmetries can be found at intermediate and larger
scale sizes (relative to the compact central disk). The change in the amplitude of the chromatic phases as a function of baseline length makes it difficult to explain this behaviour only using colour gradients, i.e. having a chromatic photocentre shift of a brightness distribution, like in the dusty region of the Circinus galaxy \citep{2013arXiv1312.4534T}. This reasoning motivates us to use asymmetric shifts to explain the behaviour of the chromatic phases on NGC 1068.

\subsection{Variability}
\label{subsec:variability}

The interferometric data of NGC 1068 was taken over a period of 7 years and there
is some evidence that the nucleus of this source is variable  \citep{1997Ap&SS.248..191G,2006A&AT...25..233T}. Therefore
before we attempt to model our data, we need to investigate whether source variability may 
influence our measurements. To this end we compare a $(u,v)$ point measured at two
different epochs. This can provide us with information about
the source evolution and/or the reliability of the  instrument itself. 
In our dataset we have a visibility point measured using 
the ATs in 2012 at a projected baseline $BL=40$ m at a position angle of 36$^{\circ}$.
This point was  measured in 2005 by \citet{2009MNRAS.394.1325R} using the
8.2-meter Unit Telescopes. Figure~\ref{Figcheck} shows the correlated fluxes and
chromatic phases of this point at both epochs.  The general trend of the spectra
are consistent with each other; we only see  some small deviations between the
$9.7-10.7$ $\mu$m  and $12.5 - 13.0$ $\mu$m, close to the regions with atmospheric
absorption.   The chromatic 
phases are mostly similar except for some small deviations around 10.5  and 12 $\mu$m. 
This $(u,v)$ point includes most of the flux of the small hot region\footnote{We refer to the small hot region as the 800K component reported by \citet{2009MNRAS.394.1325R}.}. 
We expect any variability to arise from the central accretion disk and
the effects of a change of luminosity from this heating source to 
first influence the hot dust located close to the center and only later on the more distant dust.
The flux from this component outside the silicate absorption feature is $\sim 5$ Jy
(c.f. Fig.~\ref{Figcheck} and the modelling below) and the change in this flux
as estimated from Figure 4 is $<$0.5 Jy.  So we may conclude that the mid-infrared
nuclear flux variation in this 7 year period did not exceed $\sim 10$\%.  Given this upper
limit we include all MIDI data, regardless of epoch, in our modelling.

\begin{figure*}
\centering
\begin{minipage}{.5\textwidth}
  \centering
  \includegraphics[width=\linewidth]{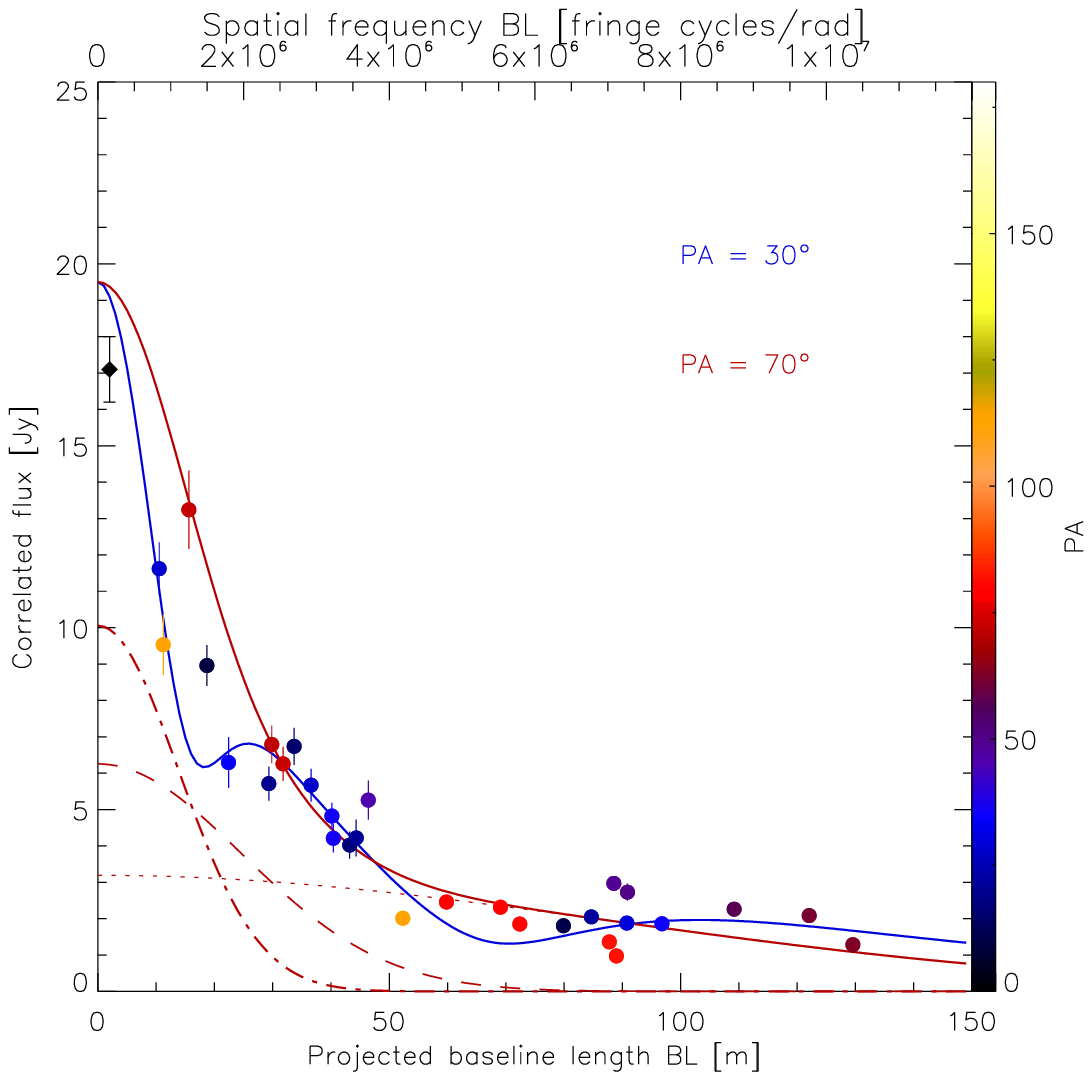}
\end{minipage}%
\begin{minipage}{.5\textwidth}
  \centering
  \includegraphics[width=\linewidth]{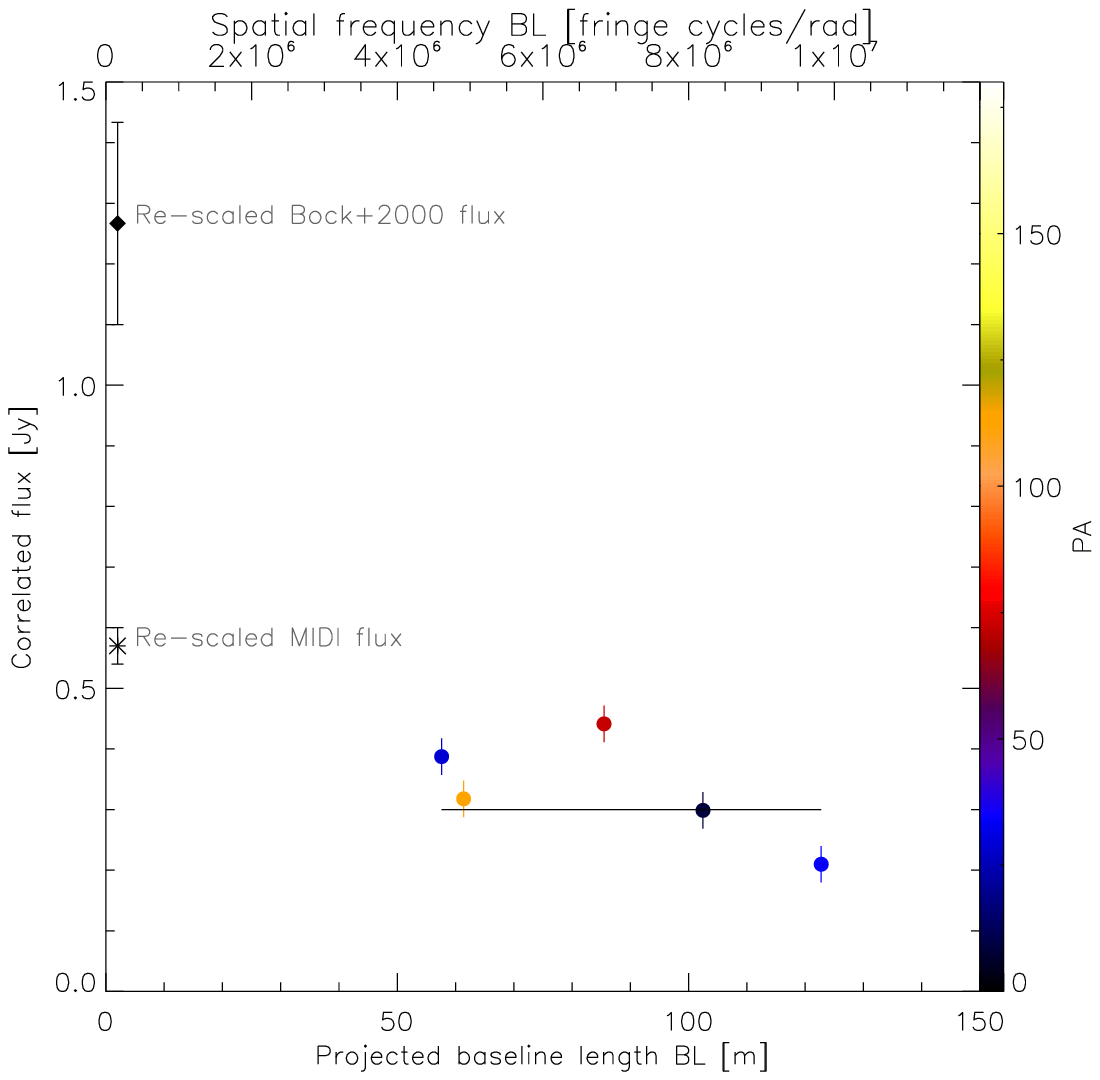}
\end{minipage}
\caption{Correlated fluxes  of NGC 1068 at $\lambda=12.5 \mu$m as a function of the projected baseline length $BL$. Left) The data is colored according to their respectively position angle. The solid lines represent the correlated fluxes obtained at two different position angles from the Gaussian modelling (model 1). The contribution of each component is represented by different lines, the dotted line represents the 1st component, the dashed line is the 2nd component and the dash-dotted line the 3rd component for a PA=70$^\circ$. Right) Expected radial plot using the photometry from \citet{2000AJ....120.2904B} for NGC 1068 if the source is 'placed' at a distance $\sqrt{30}$ times as far away as its current position. }
\label{radial12}
\end{figure*}

\subsection{Radial profile of the correlated fluxes}
\label{subsec:radial}

Figure~\ref{radial12}  shows the radial profile of the correlated fluxes at 12.5~$\mu$m as a function of the projected baseline length $BL$. The total single dish flux is obtained from the masked total flux obtained with MIDI as reported in \citet{2013A&A...558A.149B}. This flux is limited by a mask with a FWHM of $\sim$500 mas and includes the nuclear core emission. 

The correlated fluxes show a rapid drop  from values around 13 Jy at $BL\approx10$ m to less than 3 Jy at a $BL\approx50$ m. Longer projected baselines (50 m - 140 m) show an almost constant value between 1.0 Jy and 3.0 Jy. The nearly constant behaviour of the correlated flux as we go from  50 m  to 140 m projected baseline length means that the source of emission cannot be resolved and it must consist of one or more small regions of size smaller than the resolution limit of the interferometer $\lambda/(2BL)$, i.e. less than 9 mas.   On the other hand we can also set a size limit for the large scale structures that can be observed with the interferometer. Since our lowest projected baseline length is approximately 10 m, the largest structures that can still be resolved correspond to emission regions with a  diameter of $\sim$260 mas.

We observe that 14\% of the MIDI masked flux corresponds to the emission regions with a size smaller than 50 mas in diameter while the remaining 86\% corresponds to the large scale structures inside the core region with sizes between 50 mas and 500 mas in diameter, the upper limit is given by the resolution limit of the single dish telescope. 

The scatter seen in Fig.~\ref{radial12} for similar projected baseline length is caused by the position angle dependency of the correlated fluxes. Since the $(u,v)$ coverage of NGC 1068 is not equally mapped in all directions we cannot make a proper analysis to infer information about the source size in different directions. Still at the lowest projected baseline lengths we can find three $(u,v)$ points with a similar length  (BL = 10, 11, 15 m) but observed with different position angles (at 28$^\circ$, 72$^\circ$ and 113$^\circ$ respectively). In fact the point with the longest projected baseline length at PA= 72$^\circ$ has the highest correlated flux. This indicates that the source is less resolved in this direction than the others.

\subsection{NGC 1068 in the context of the Large Program (LP) study}

Results from a large survey of AGNs using  mid-Infrared interferometry to resolve the small scale structures of their nuclear regions  were presented in \citet{2013A&A...558A.149B}.  NGC 1068 and Circinus are a factor of 10 brighter than the rest of the sources analyzed in that work and they seem to have significant differences compared to the rest of the sample. A re-scaling of these two bright sources was performed in \citet{2013A&A...558A.149B} to observe the radial behaviour of the correlated fluxes with a resolution and flux similar to the more distant, weak sources.  In the case of NGC 1068 the source was ''placed" at a distance $\sqrt{30}$ times as far away as its current location so that its fluxes would match the median flux of the weak sources.

With our new short baseline observations we can more accurately repeat the rescaling experiment and give a quantitative description of the resulting properties when observed under conditions similar to the weak targets. To make this experiment as realistic as possible we look only at the rescaled baseline lengths which are available with the UTs. Figure~\ref{radial12} shows the radial plot of NGC 1068 after applying the rescaling at $\lambda=12.5 \mu$m.

To completely match the conditions of the observations of NGC 1068 with the the weak sources we have to take a photometric flux with a similar resolution as the rest of the targets. For some of the  weak sources the total flux was extracted  from  MIDI single-dish observations using a window of size 0.''5 x 0.''52  or by taking high spatial resolution spectra or photometry obtained from the VLT spectrometer and imager for the mid-infrared (VISIR) using a 0.''53 window in the spatial direction. If we re-scale NGC 1068 to a distance which is  $\sqrt{30}$ times further away, then the 0.''53 window corresponds to a $\sim$ 3'' window in the normal space. From \citet{2000AJ....120.2904B} we know that the flux  at 12.5$\mu$m does not vary too much  from a beam of a diameter of 4'' to a beam of a diameter of 2'', since we need a beam of diameter of 3'' we take the value of the 4'' diameter beam, $F_\textrm{Bock}=36$ Jy scaled to 1.2 Jy, to be a reasonable approximation. Although the results from \citet{
2013A&A...558A.149B} were obtained at $\lambda=12 \mu$m and we are going to use $\lambda=12.5 \mu$m for our experiment\footnote{Bock et al. (2000) only reports values at $\lambda=7.9,10.3,12.5,24.5\mu$m. This is the reason why we perform our analysis at $\lambda=12.5\mu$m.}, the results do not vary within this wavelength range since the emission is mostly determined by the continuum and no absorption effects come in to play.

Fitting a radial model to the rescaled data, following\citet{2013A&A...558A.149B}, 
by using a two-component model with an unresolved ''point source''  and a Gaussian component we
find that the point source fraction ($f_p$), of the total flux is of the order of 24\% and the FWHM of the Gaussian must be $\gtrsim$36 mas (the data only provide a lower limit to the size). At the rescaled distance the rescaled limit of the Gaussian component corresponds to structures larger than 14 pc. 
Figure~\ref{radial12} shows that the flux that we attribute to a "point source'' comes from a 
partially resolved structure showing a position angle dependence 
which could be attributed to an elongation.  

\citet{2013A&A...558A.149B} classified NGC 1068 as consisting of  a resolved plus  an unresolved emitter. 
In our rescaling experiment, where we match the fluxes and conditions of observation to the other weak sources, we now classify the rescaled version of  NGC 1068 as an unresolved plus an overresolved emitter. 
We can now compare the results of our experiment with the results of \citet{2013A&A...558A.149B} (c.f. Figs. 30, 31, 32, and 34), where the principle difference with the earlier work is that $f_p$
has increased from 0.1 to 0.24.  This ''rescaled" value of $f_p$ is no longer detached 
from the sample distribution, but appears at the lower edge, similar to Seyfert 
galaxies NGC 4507 and MCG-5-23-16.  Thus in the context of the survey, NGC~1068 has a 
large but not extraordinary flux fraction in well-resolved structures.

Our experiment shows that the difference between the value of $f_p$ of NGC 1068  
and the typical values for Seyfert Type 2 sources in \citet{2013A&A...558A.149B} can be in part
attributed to resolution effects. This becomes more evident in a plot of the $f_p$ versus intrinsic resolution
(see Fig. 32 in \citet{2013A&A...558A.149B}). However, the new rescaling experiment
does not contradict the conclustion of \citet{2013A&A...558A.149B} that the spread in
morpholgies observed in the survey is instrinsic and not a resolution effect,
NGC~1068 still shows some structural differences
from the rest of the LP sample.  It shows significantly non-zero
chromatic phases at baselines of  10-20 meter.  These correspond to baselines
of 60-120 meters at the ``rescaled" distance.  None of the more distant LP galaxies
show non-zero chromatic phases at these baselines.  Circinus seems to resemble
NGC~1068 in this respect.

\section{Gaussian fits}
\label{sec:gaussfits}

\subsection{Motivation}

Direct image reconstruction techniques cannot be applied to our interferometric data
for two reasons: the limited $(u,v)$ coverage and the lack of true phases. 
Still, information can be recovered by using   simple
analytical forms, such as gaussians or point sources, to describe the source
brightness distribution. The reason for using this approach is to describe  the
brightness distribution as accurately as possible  with a small number
of parameters and making few specific physical assumptions. 
In this section we present model fits for the recent
observations, that allow us to specify the mid-infrared geometry of 
the  1-10 pc region of  NGC 1068.

\subsection{Greybody gaussian models}
\label{subsec:gaussmodels}

Greybody gaussian models have been used in earlier papers to model the amplitudes of
the correlated flux  and get estimates of the sizes,  temperature and inclinations
for components in different AGN (e.g. NGC 1068: \citet{2004Natur.429...47J, 2009MNRAS.394.1325R};
Circinus; \citet{2007A&A...474..837T,2012JPhCS.372a2035T}; Centaurus A:
\citet{2009ApJ...705L..53B}. So far the modelling only included the amplitude of the
correlated flux and thus by necessity assumed that the gaussian components were concentric and
therefore symmetric. 

\citet{2009MNRAS.394.1325R} showed that the hot small emission region can be
described by a gaussian grey body with an absorption screen in front that reproduces
the silicate feature. The authors described the larger scale emission region 
with a second component, but not enough short baseline information  was available 
to constrain the parameters correctly. This concentric two-component model 
agreed well with the amplitudes of the correlated fluxes but by definition cannot
reproduce the non-zero chromatic phases described here.

For this work we again use a multi-component  greybody gaussian model to fit the
mid-infrared interferometric observations of NGC 1068. The model treats the infrared
emission as coming from gaussian grey body components of a fixed size temperature and
orientation, each one behind a uniform absorption screen, but we no longer
require the components to be concentric.  The contribution to the
correlated flux of each component  for a $(u,v)$ point is given by
\begin{equation}
 F_{corr}^i(\lambda,u,v)  = \eta_i BB_\lambda (T_i) \frac{\pi \Theta_{i}
\theta_{i}  }{4 \ln 2 } V_i(u/\lambda,v/\lambda) e^{-\tau_i C_{abs}^i(\lambda)}
\end{equation}
where $T_i$, $\tau_i$, $\eta_i$ are the dust temperature, the optical depth and the
scaling factor of the  component $i$, respectively.  The scaling factor  $\eta_i$ can
be considered to be a surface filling or emissivity factor, which is independent of the wavelength and limited to values of $0<\eta\leq 1$. $V_i$
 is the contribution of the visibility of an elongated gaussian component, described
by a FWHM ($\Theta_{i}$) along the  major axis,  a FWHM  ($\theta_{i}$) along the
minor axis and a position angle ($\psi$). The visibility function $V_i(u/\lambda,v/\lambda)$ is obtained from computing the Fourier transform of a Gaussian intensity distribution function described as,

\begin{equation}
 G_i(\alpha,\delta,\psi)  =  \exp\left\{-4 \ln 2  \left[ \left(\frac{\alpha_i'}{\theta_{i}}\right) + \left(\frac{\delta_i'}{\Theta_{i}}\right) \right] \right\}
\end{equation}
where $\alpha_i'=(\alpha-\alpha_i)\cos\psi_i-(\delta-\delta_i)\sin\psi_i$ and  $\delta_i'=(\alpha-\alpha_i)\sin\psi_i+(\delta-\delta_i)\cos\psi_i$ are the positional coordinates of the Gaussians. ($\alpha_i,\delta_i$) is the center of the $i$th Gaussian component.
The absorption curve for the
chemical composition associated with the i-component is described by $C_{abs}^i$. For the dust absorption curves
we have selected 3 dust absorption templates, including Ca$_2$Al$_2$SiO$_7$
(gehlenite, \citet{1998A&A...333..188M} ) which was found as the best fit in
\citet{2004Natur.429...47J}, the standard galactic dust as observed towards the
center of our Galaxy \citep{2004ApJ...609..826K} and $\alpha$-SiC
\citep{1993ApJ...402..441L} suggested by \citet{2010MNRAS.406L...6K} as a better
explanation for the anomalous absorption feature present in NGC 1068. 
The dust template used for each component is made from a linear combination of the
three mentioned dust templates. The coefficients for each component are fitted
along with the rest of the parameters. 

The final form of the complex correlated flux will be given by,
\begin{equation}
 F(\lambda,u,v)= \Sigma_j F_{corr}^j(\lambda,u,v) e^{-2\pi i(u_\lambda\cdot \alpha_j+v_\lambda\cdot
\beta_j)}  
\end{equation}

where $\alpha_j$ and $\beta_j$ are the offset in right ascension and declination,
respectively and $u_\lambda=u/\lambda$ and $v_\lambda=v/\lambda$. The total single dish flux can be recovered by using the coordinates $u=0$ and $v=0$, i.e.  $BL=0$ m.

\subsection{Offset components}

\begin{table*}
\caption{Parameters found for the greybody gaussian models. See Sect.~\ref{subsec:gaussmodels} for a
description of the parameters.}             
\label{table:1} 
\centering     
\scriptsize
\begin{tabular*}{\textwidth}{c | c | c c c c c c c c c  | c }        
\hline\hline  
\multicolumn{12}{c}{\citet{2009MNRAS.394.1325R}. Amplitudes of UT data} \T \\
\hline
Model & $i$ & $T$ & $\tau$ & $\eta$ & FWHM major  & FWHM minor  & P.A. & $\alpha$ &
$\beta$ & Dust template & Reduced  \T \\    
 & & [K] &  &  & [ mas ] ( [ pc ] ) & [ mas ] ( [ pc ] ) & [degree] & [ mas ] ( [ pc ] ) & [
mas ] ( [ pc ] ) & & $\chi^2$  \B \\
\hline                        
 &1 & 800   & 1.9 & 0.25 & 20 ( 1.4 ) & 6.4  ( 0.4 ) & -42 & 0 & 0 & Fitted
composite & \T \\
 0&2 & 290   & 0.42 & 0.64 & 56.5  ( 3.9 ) & 42.4 ( 2.9 ) & 0 & 0 & 0 & Fitted
composite &  - \\
 &3 & -   & - & - & -  ( - ) & - ( - ) & - & - & - & - & \\
 \hline\hline 
\multicolumn{11}{c}{Data used: Amplitudes of UT + AT data and chromatic phases of AT
data} \T \\
\hline
 &1 & 660 $\pm 26$& 1.6 $\pm 0.2$ & 0.31 $\pm 0.03$ & 20.9 $\pm 1.7 $ ( 1.46 ) &
7 $\pm 0.8$ ( 0.49 ) & -45 $\pm 4$ & 0 & 0  & 50\% Gehlenite + & \T  \\
 & & \multicolumn{8}{c}{} & 50\% Galactic dust &  \\
 1&2 & 257 $\pm 15$ & 0.98 $\pm 0.4 $& 1 $^{+0}_{-0.23}$ & 53 $\pm 8.5$ ( 3.7 ) & 23
$\pm 6.8$ ( 1.6 ) & 120  $^{+15}_{-30}$ & 0 $\pm 8$ ( 0 ) & 19$\pm 4 $ ( 1.32 ) & Galactic
dust & 6.24 \\
 &3 & 360 $\pm 36 $ & 0 $\pm 0.3 $ & 0.047 $\pm 0.016$ & 185 $^{+90}_{-50}$ ( 12.9 ) &
50 $^{+60}_{-20}$ ( 3.5 ) & -36 $\pm 7$ & -30 $\pm 17$ ( -2.1 )& 100 $\pm 12$ ( 7 )&  Galactic dust
& \T \\
 \hline
 &1 & 700 $\pm 30$& 1.25 $\pm 0.15 $ & 0.23 $\pm 0.02 $ & 20 $\pm 1.4$ ( 1.4 ) &
6.8  $\pm 0.7 $ ( 0.47 ) & -45 $\pm 4 $ & 0 & 0  & 50\% Gehlenite + & \T  \\
 & & \multicolumn{8}{c}{} & 50\% Galactic dust &  \\
 2&2 & 301 $\pm 21 $& 2.8 $\pm 0.8 $ & 1 $^{+0}_{-0.2}$ & 42 $\pm 9 $ ( 2.9 ) & 
29 $\pm  8$ ( 2 ) & 102 $\pm 36 $  & 3.7 $\pm 8 $ ( 0.25 ) & -17.4 $\pm 7 $ ( -1.2 ) & Galactic
dust & 6.23 \\
 &3 & 370  $\pm 35 $& 0.15 $\pm 0.3 $ & 0.05 $\pm 0.017 $ & 200 $^{+80}_{-60}$ ( 14
) & 52 $^{+65}_{-22}$ ( 3.6 ) & -35 $\pm 6 $  & -24 $\pm  14 $ ( -1.7 ) & 80 $\pm 10 $ ( 5.6 )& 
Galactic dust &  \B \\
 \hline
\end{tabular*}
\end{table*}

In addition to the amplitude of the correlated fluxes we have also measured
the chromatic phases. As already mentioned in Sect.~\ref{subsec:chromaticphases} these phases
are not the true phases, but still provide some spatial constraints.  
The properties of the chromatic phases observed in NGC 1068  and their distribution on in the $(u,v)$ plane excludes the possibility
that all major components are concentric.  Non-concentric components are required to reproduce
the asymmetries observed, which are present for a large range of projected baselines
in our ATs observations and also a few in longer baselines (See Appendix for plots
of the UT data).  The longest UT baselines, which respond primarily to the emission
from the hot core, show small phases, indicating that this core is essentially
symmetric.  The intermediate and short baselines respond to both the core and
the larger structures and in the current models the non-zero phases indicate
displacements between these components.

We can use equation 2 to calculate the complex correlated flux  not only for the
computation of the amplitude, but also for the calculation of the  chromatic phases.
For each modelled visibility point the  constant and linear dependencies on wavenumber
are removed from the complex correlated flux. By doing this we lose some
information of the exact position of each component in the real space, but the
information about the positions with respect to each other is preserved. To
reduce the degeneracy of solutions we have fixed the position of the smallest
component to be at the center of our plane, i.e.,  $\alpha_1=0$ and $\beta_1=0$.

We have spectra of the amplitudes and chromatic phases 
for 30 $(u,v)$ points, which include information from projected baselines between 10m and 130m. 
We have attempted to fit 3-component models to these data, but
we have not been able to find a reasonable fit to all the data. 
In these attempts we can find  reasonable fits for either the 
chromatic phases or the correlated fluxes but not both together. 
Apparently a more complicated model is necessary.  Because
our immediate goal was to describe the emission morphology
with a limited number of parameters, we have chosen
not to add a fourth or further components to improve the fit.

Instead we will focus our attention on describing only the spatial information 
contained in the AT observations. This approach ignores the details
of the small scale structures but describes the relative positions of the three components with respect to each other. The result from this procedure can be interpreted as
a low resolution image of the emission of NGC 1068.  It is our intention
to present at a later date representations of our data using 
more general image reconstruction techniques constrained by physically
motivated emission models.

The parameters reported in this paper  
are the least-squares solution found using the Levenberg-Marquardt
technique to fit  all 30 amplitude spectra  and the 11 chromatic phase spectra from
AT observations reported here. The reason of not using the chromatic phases measured with the UTs is to give an equal weight to the chromatic phases in all the scales, otherwise the fitting routine would focus more in the intermediate scale regime and the aim of this work is to disentangle the large structures from the small scales.  
Additionaly, we do not try to fit the single dish MIDI masked flux  since it was measured using the UTs. As mentioned in Sect.~\ref{subsec:radial}, the masked flux could also be capturing part of emission comming from regions further than 250 mas in distance. 

There is a difference in the field of view of the UTs and ATs, which  means that the region observed by the ATs might include regions which are not captured by the UTs. This does not seriously affect the interferometric observations since the UTs observe the small scales and in Sect.~\ref{subsec:variability} we 
showed that the UTs and ATs observe the same structures on the small size regime (>30 mas). The large scale structures are over-resolved with the UT baselines and thus depend only on the information obtained from the AT baselines. 

To estimate the errors in the parameters, we  fixed the value of 
all but one parameter at a time. The uncertainties that we report represent the range of 
variation of each parameter sufficient to cause a significant change to the quality of fit. 
We saw that parameters with a reduced chi-square larger than 15\% of our best reduced chi-square begin to show significant differences.

We first attempted to fit a two component model as
\citet{2009MNRAS.394.1325R} but relaxing their assumption of concentricity.
We were unable to get a reasonable fit of our data.  We then tried three component models.

\begin{figure*}
\centering
\begin{minipage}{.5\textwidth}
  \centering
  \includegraphics[angle=0,width=\hsize]{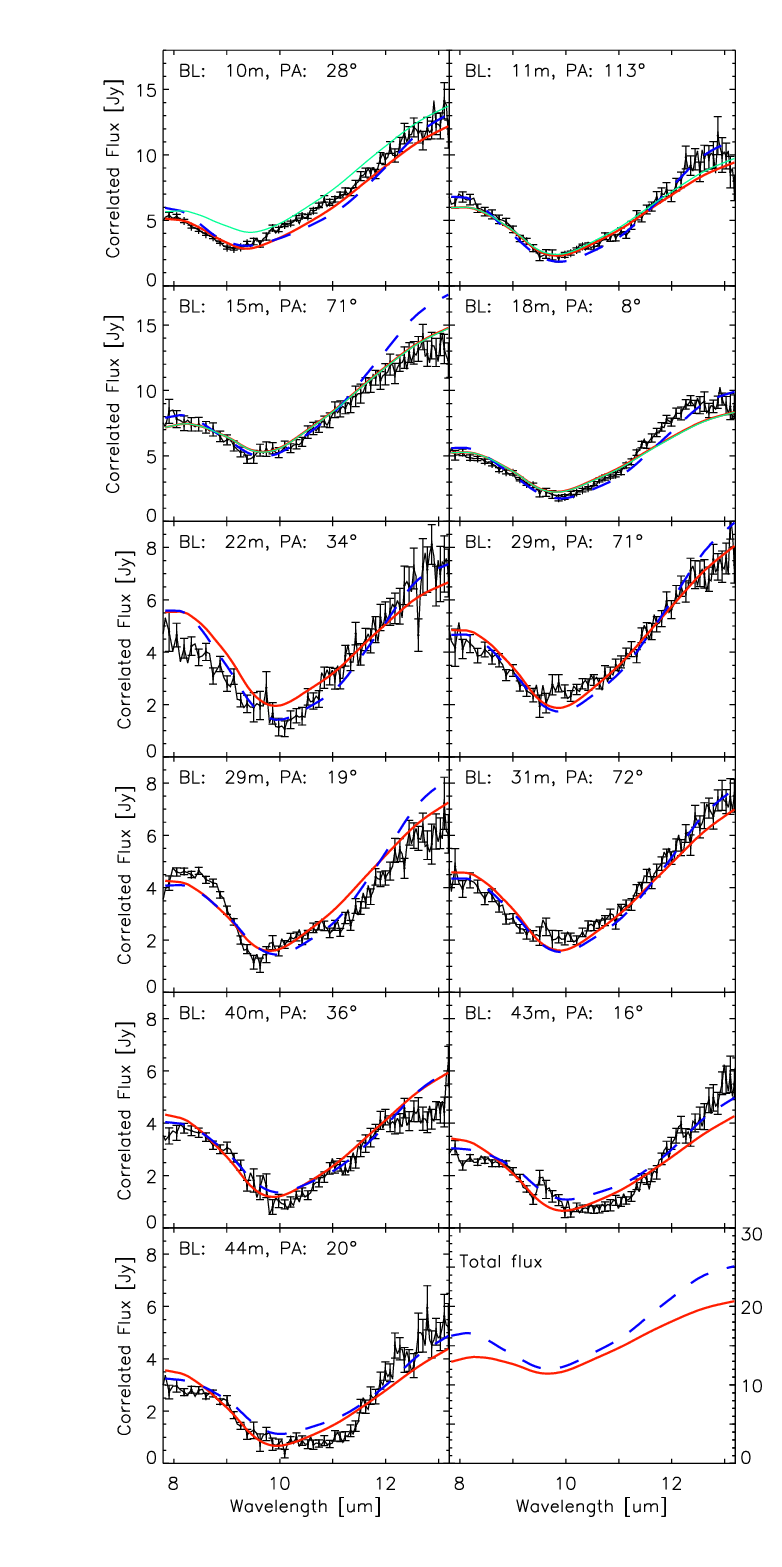}

\end{minipage}%
\begin{minipage}{.5\textwidth}
  \centering
  \includegraphics[angle=0,width=\hsize]{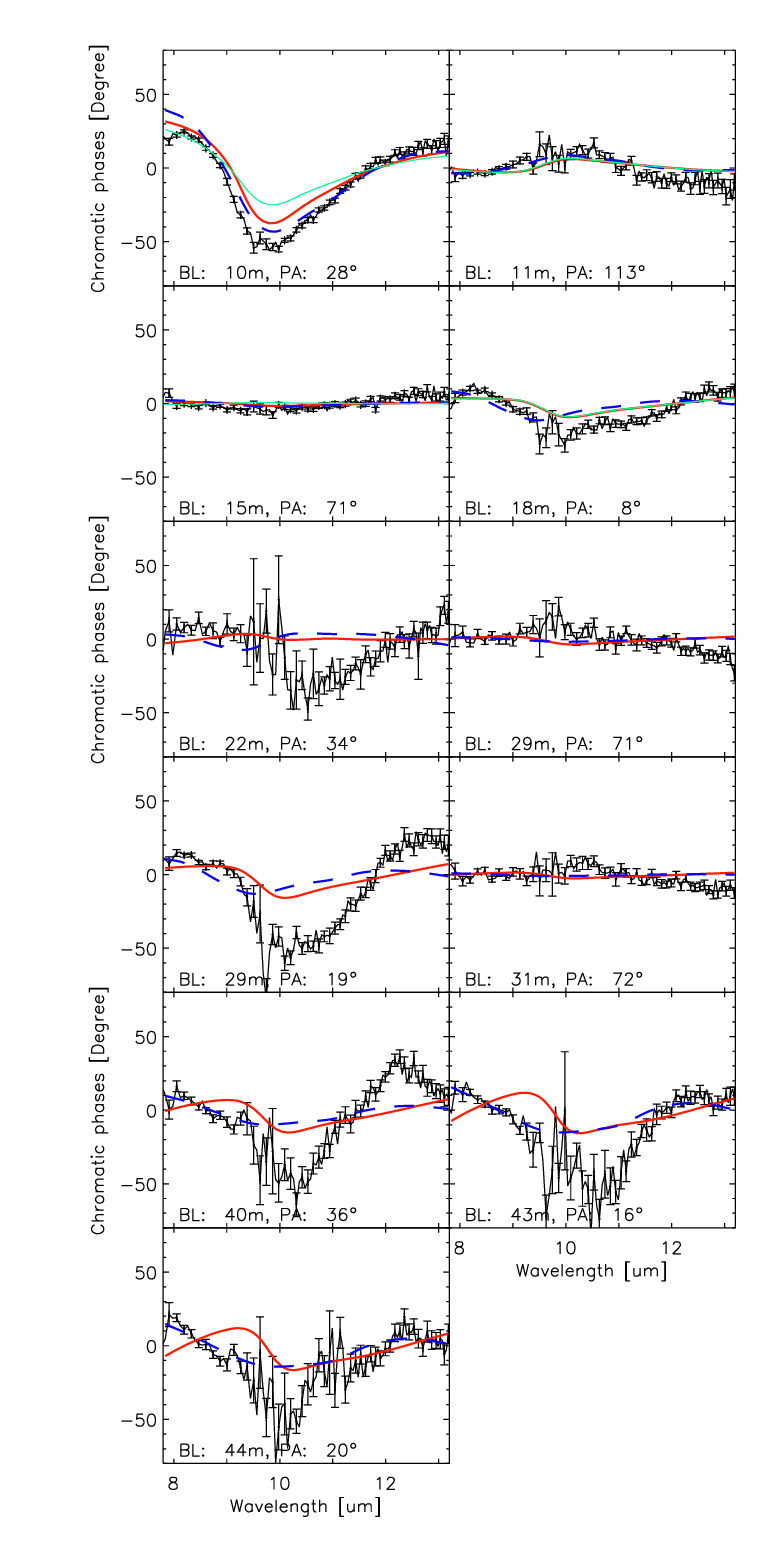}
\end{minipage}
\caption{Plots showing the best fit model for the $(u,v)$ points observed with the ATs
Left) Correlated fluxes and Right) Chromatic phases. The red line shows the
corresponding correlated fluxes and chromatic phases for our 1st best non-concentric model while the blue dashed line represent our 2nd best non-concentric model. Light green lines (only on the first 4 plots of each panel) show the curves obtained by taking  the parameters of our 1st best model and taking the lower limit of the uncertainty in the offset declination for the 3rd component, we observe some significant deviations in some of the plots.}
\label{Figfit}
\end{figure*}
   
We found 2 sets of parameters which can describe the source brightness distribution
and the chromatic phases to a reasonable degree; these two sets of parameters  can
be seen in Table 1. Both models consist of two  warm components using a standard
galactic dust template  and one hot small component using a mixture of gehlenite and
galactic dust as the dust template. Contrary to what  \citet{2010MNRAS.406L...6K}
report from their study, we find that the contribution of SiC in the dust templates
is at most very small.  \citet{2010MNRAS.406L...6K} used the spectra of  only one $(u,v)$ point in
their modelling, and the optical depth of the SiC feature can be chosen
to fit well this one point, but cannot fit the ensemble of observations.
In figures~\ref{corrut} and ~\ref{phiut} of the Appendix we plot also
best fit curves obtained for a model using only SiC as the template.
The poor match to most of the data indicates why the least-square routine
avoided this component to the fits. 

The sizes and position angle of the hot component for both models agree  with the
parameters previously reported by \citet{2009MNRAS.394.1325R}. This is not
surprising since we left out the chromatic phases measured with the UTs, i.e. long
projected baseline, therefore the hot small emission region will be treated as a
symmetric gaussian component. The 660 K temperature for one of our models is clearly
lower than the one reported by \citet{2009MNRAS.394.1325R}, the differences might be
attributed to the dust absorption templates used in each model.

\begin{figure*}
\centering
\begin{minipage}{.3\hsize}
  \includegraphics[width=\hsize]{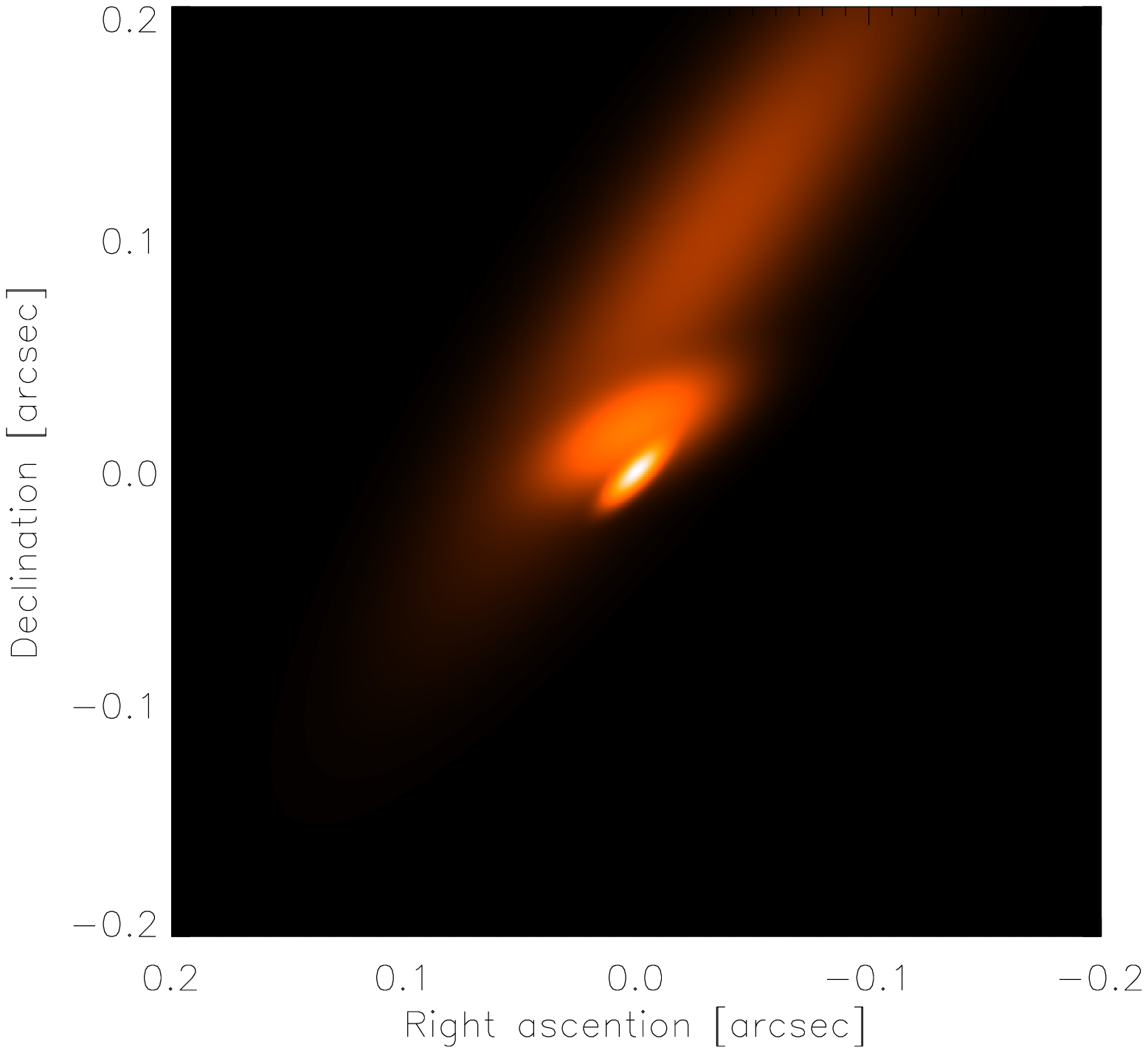}

\end{minipage}%
\begin{minipage}{.38\hsize}
  \includegraphics[width=\hsize]{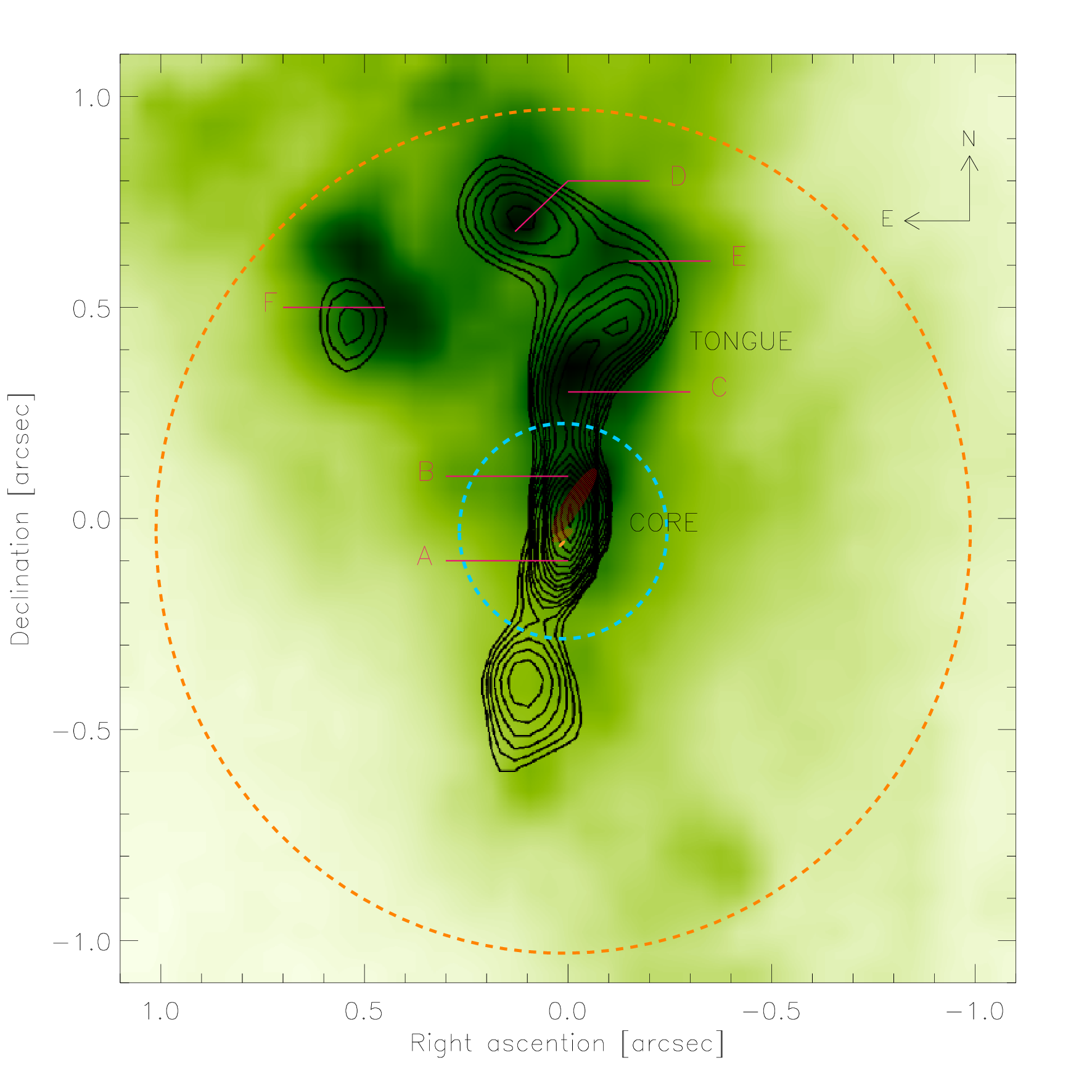}
\end{minipage}
\begin{minipage}{.3\hsize}
  \includegraphics[width=\hsize]{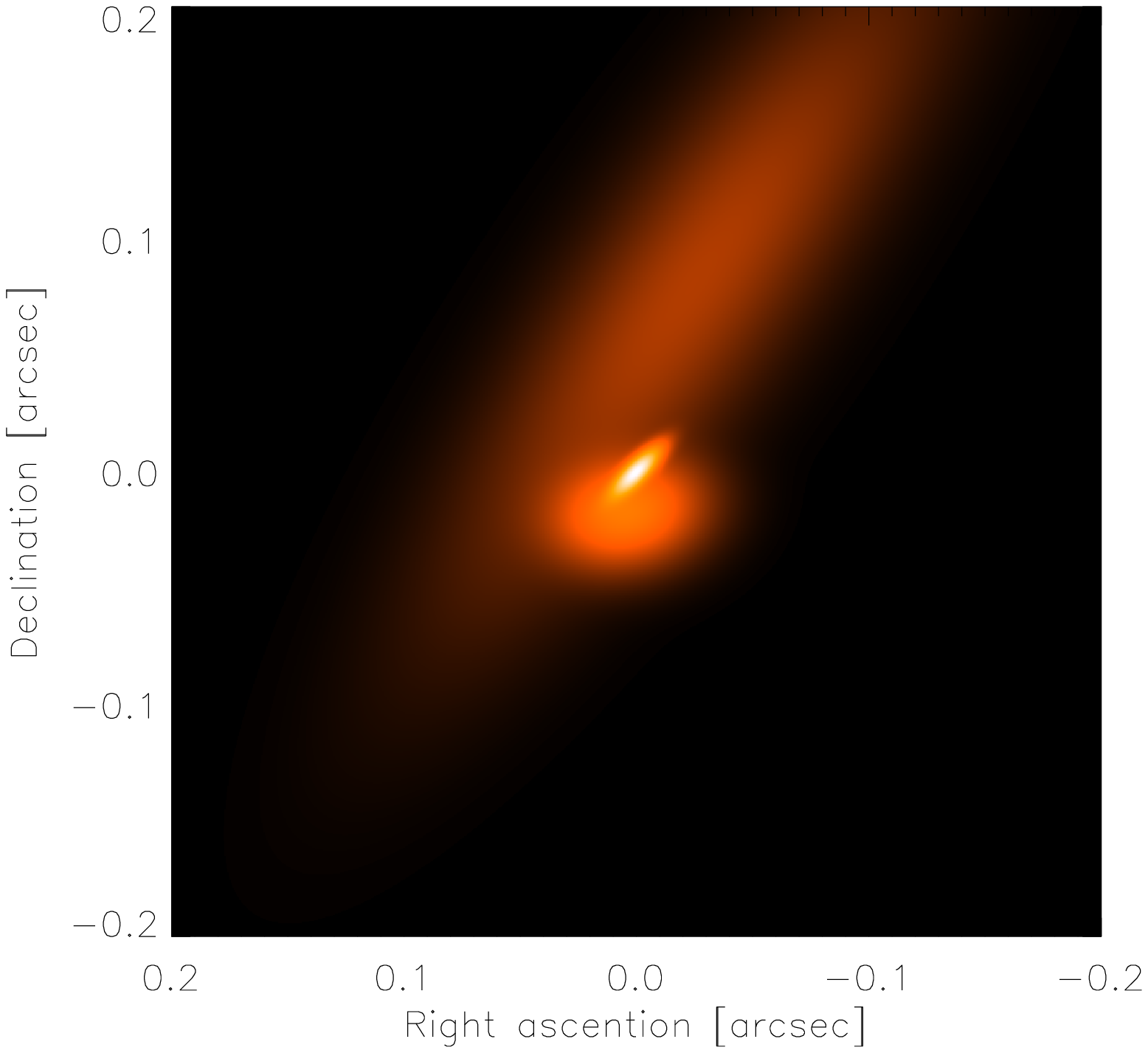}

\end{minipage}%

\caption{Image of the No. 1 ({\it Left}) and No 2. ({\it Right}) best three component models for the mid-infrared emission at 12.0 $\mu$m of the nuclear region of NGC 1068. The image was scaled using the square root of the brightness.   {\it Center)} Comparison between our
first best model and the 12.5 $\mu$m image of \citet{2000AJ....120.2904B}, taken with the 10m Keck telescope. The dashed circles represent the FWHM of the field of view for MIDI using the UTs (blue) or the ATs (orange). The letters indicate the positions of the [OIII] clouds according to \citet{1991ApJ...369L..27E}}
\label{sketch}
\end{figure*}

The two best fitting models differ primarily in the position of the
intermediate sized component.  In our first model (c.f.  Fig.~\ref{sketch})
we find this component to be associated with emission of  dust around 257 K  with a size 
of 3.7 x 1.6 pc elongated at a position angle of 120$^\circ$. 
The position of its center  is 19 mas to the north of the hot component. 
The third component is a large emission region with
a temperature around 360 K.  The center of the Gaussian which
represents this large region is $\sim $ 100 mas  to the north of the hot component and at an angle of -18$^\circ$ (measured from North to East). It is a highly elongated structure with an axis ratio
of $\sim 1:3.7$ and a position angle of  $-36^{\circ}$. 

Our second best fit model also has an intermediate and large size gaussian component. The
second component for this model consist of emission at 300 K  with an intermediate
size of 2.9 x 2 pc elongated at a PA of 102$^\circ$ and located 17 mas south
of the hot component.  It is absorbed by a screen with a
larger  optical depth close to 3. The 3rd and biggest component with temperature
around the 370 K has a FWHM of 200 mas along the major axis and a FWHM of 52 along
the minor axis. The position angle of this large component is similar to the one
for our first model, $-35^{\circ}$. The center of this large emission region is
also located to the north-west of the small hot component but with a slightly
different position than our first model, it can be found at a  distance of $\sim $ 80
mas  from the center and 16$^\circ$ in the NW direction. The ratio between the major
and minor axis is also close to $\sim 1:4$.

The main differences between our best fit models lie in the parameters of the
intermediate size component. In the second model this component lies to the south rather
than to the north, and it has a much larger absorption optical depth.  The degeneracy
of the modelling, i.e. the existence of two equally good fits is caused by the
limitations of using chromatic phases.  The contribution of this component to the
phases changes sign when it is moved to the south,  but this effect is cancelled,
after removal of the linear phase gradient, by the absorption-diminished emission of
this component near 10$\mu$m.  This ambiguity would be removed if true phase or
closure phase observations were available.  These possibilities should be available
with future VLTI instruments, e.g. MATISSE \citep{2008SPIE.7013E..70L}.
 
The fitted correlated fluxes and the chromatic phases for both models can be seen in
Fig.~\ref{Figfit}. The correlated fluxes fit  well;
deviations are mainly caused by the assumption of a gaussian shape.   The largest
deviations are seen in the mid- and long- baseline range where the resolved structures
are probably  more complex. This is also seen in the chromatic phases; the shortest
baselines are fitted well with our model. In the long baseline regime (see
Online Appendix for plots of UT chromatic phases) the chromatic phases are mostly
zero and thus our model is consistent in that regime but in  the intermediate regime
the model fits poorly.  

\section{Discussion}
\label{sec:discussion}

\subsection{Summary of modelling}

Our three Greybody models provide us with a general image of the main emission
regions of NGC 1068 using information from the short baselines and
chromatic phases. The main differences between our models and the work by \citet{2009MNRAS.394.1325R} are the new short spatial frequency observations and the use of chromatic phases to get  spatial information of the nuclear mid-infrared region. A sketch of the emission region can be
seen in Fig.~\ref{sketch}. Although our models
are not able to reproduce the chromatic phases completely, especially in the
intermediate baseline regime, we observe a reasonable consistency in the lower and
higher baseline regime which fit much better than purely concentric models. 
 The poor fits observed in the intermediate baseline regime
would suggest that the link between the small scales and the large scales is more
complex than can be fitted with a limited number of gaussians
and has to be modelled more carefully in order to understand the relations between the
various size scales. Still, our 3-gaussian model represents reasonably well the global variation of surface brightness with position in the nuclear region of NGC~1068.  The resolution of these variations into discrete components may be an artefact of the modelling.  More sophisticated image reconstruction techniques may remove these artificial transitions, but reliable multiwavelength image reconstruction algorithms that can accommodate the chromatic phase information are not yet available.

\begin{figure}
   \centering
   \includegraphics[width=\hsize]{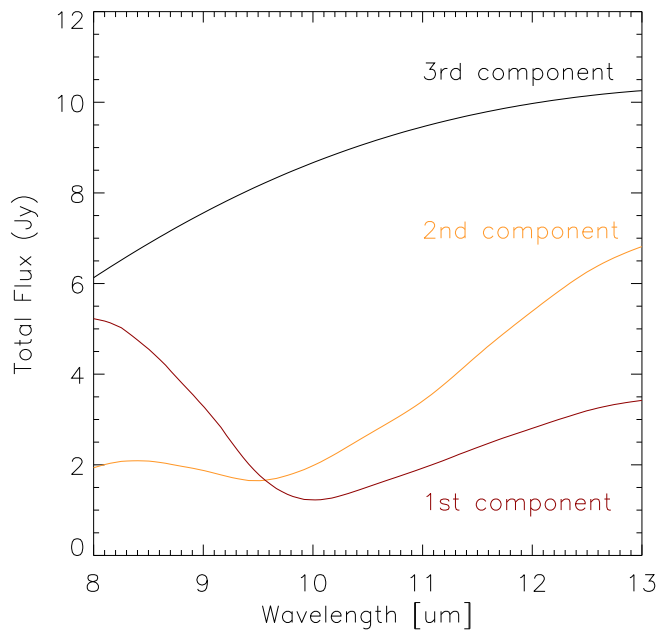}
      \caption{
      Total flux of the three best-fit model components as a function of wavelength.
              }
         \label{percentage}
   \end{figure}

Both models show that the maximum separation, in the north-south direction, between the center of the Gaussian components must be approximately 100 mas. Since we identify the component at ($0,0$)  with the radiation from near-nuclear dust (See Sect.~\ref{subsec:identification}), 
the large flux from the northern offset component must come from the regions located near 
the narrow line ionization region.  
This northern component is large relative to the other components in 
both models and the absorption optical depth and scaling factor are  small. The total flux of this component reaches 10 Jy at 12 $\mu$m. Fig.~\ref{percentage} shows the total flux of each component as a function of wavelength, the ratio of the total fluxes of each component varies depending on the wavelength. We observe a ratio of the fluxes of 1 : 0.4 : 1.2 for the 1st, 2nd and 3rd component respectively at 8 $\mu$m, 1 : 1.6 : 7 at 10 $\mu$m and 1 : 2 : 3.5 at 12 $\mu$m. Thus our 3rd (the largest component) contributes with 46\%, 73\% and 55\% of the total flux of structures with sizes below 260 mas in diameter, i.e. the ones that are resolved with the interferometer\footnote{This information was extracted using our model no. 1. Similar results are obtained for the model no. 2}. 

The position and absorption profile of the intermediate component are ambiguous in our models.  The foreground absorption 
required in the second model, ($\tau\sim 3$), is larger than that
in front of the central component.  This is counter-intuitive, but cannot be excluded
in models where the dust is distributed in irregular clumps \citep{2008ApJ...685..147N,2008A&A...482...67S}. 
This ambiguity can be removed if additional $(u,v)$ coverage at intermediate baselines
is obtained.
Still, it is interesting to see  that when comparing our two best models, the general trend seems to be consistent with a gradient in the silicate absorption, which decreases when going from  south to  north. \citet{2013arXiv1312.4534T} used this gradient in the silicate absorption to explain the behaviour of the phases observed in the Circinus galaxy in the low spatial frequency regime. In NGC 1068, the lowest spatial frequency regime seems also to be consistent with a gradient in the silicate absorption while the chromatic phases observed in the intermediate spatial frequency regime are caused by offset components that together form  a non-gaussian shape region.  

We conclude this summary by reiterating that the resolution of the emission from the galaxy into three distinct components,
as seen in Fig.~\ref{sketch}, is an approximation to the actual brightness distribution of the mid-infrared nuclear region of NGC~1068. If the {\it a priori} assumption of gaussian components is relaxed, the components
may blend into one continuous feature. The brightness distribution function for  model No. 1 can be approximately represented by a continuous distribution where the brightness decreases as $r^{-\gamma}$, with $r$ being the distance from the center of the 1st component  and $\gamma\approx1$.

\subsection{The Tongue}

Single dish mid-infrared images of NGC~1068 display a primarily N-S elongation dominated by
a specific feature which \citet{2000AJ....120.2904B} have named as the 'Tongue'. This coincides
with the regions IR-1/1b seen by \citet{2006A&A...446..813G} in short wavelength infrared single
dish  images, and component C in the VLBA radio maps of \citet{2004ApJ...613..794G}.
This region extends to the north of the core, bends to the west about 0.2" above the core and then to 
the east.   It seems to be associated with part of the [OIII] emission \citep{1991ApJ...369L..27E} and the radio continuum emission \citep{1996ApJ...464..198G}. According to \citet{2005MNRAS.363L...1G}, 
the Tongue region (identified as the NE1) 
has a flux of 11.2$\pm 2$ Jy at 12.8 $\mu$m and is 
thus the second brightest mid-infrared region in NGC~1068 after the core region. Because at least
part of this feature lies within the AT/VLTI field of view, we need to explore its relationship
to our measurements and models.  The 1.8 m diameter ATs have a field of view (FOV) of radius $\sim 1.14 $ arcsec at 12 $\mu$m  while the 8m UTs have a FOV of $\sim 250$ mas.  From the images in \citet{2005MNRAS.363L...1G} and \citet{2007A&A...472..823P} we
estimate the NE1 component to be 400 - 500 mas from the core.
The existence of NE1 within our FOV raises two questions: (1) is component 3 in our model fits in fact
identical with NE1, but incorrectly positioned due to our limited $(u,v)$ coverage and (2) even if
component 3 is distinct from NE1, do our observations place useful constraints on the morphology
of NE1? 

We have investigated whether positions of component 3 near NE1 are consistent with our AT data.
There are in fact such solutions, but we have discarded them as unphysical because (1)
the 3rd component then requires very large temperatures ($> 800 $K) to fit
the short baseline spectra and (2) the total flux of such a 3rd component exceeds 
the values reported by \citet{2005MNRAS.363L...1G} and also the ones reported by \citet{2008A&A...481..305P} at the two closest northern quadrants (3.4 Jy, 4.8 Jy and 7.5 Jy at  9.0 $\mu$m, 10.8  $\mu$m and 12.8 $\mu$m for their 1NO region and 1.2 Jy, 1.8 Jy and 3.1 Jy at  the same wavelengths for the 2NO region). 
Evidence in favor of the  existence of a near-in 3rd component is that the first two
interferometric components do not provide the large flux (25 Jy) measured by these
authors inside the 0.6'' diameter central aperture. 

We conclude that a 3rd component $<100$ mas from the core is necessary to fit the spectra at the shortest projected baselines. We now investigate whether if a new extra component at the position of the Tongue could improve the fits.  To avoid adding an excessive number of degrees of freedom to our model we have fixed {\it a priori}
several of gaussian parameters of the fourth component.
From the flux values reported by \citet{2008A&A...481..305P} we think it is reasonable to fix the
temperature of this component to 300 K with a very small optical depth and place it 400 mas to the north 
of the hot core.  From the mid-infrared images reported by \citet{2000AJ....120.2904B} we fix the
PA of the emission region to $\sim$ -40$^\circ$. We allow the modelling routines to fit the major and
minor axes, and adjust the scale factor of this component to fit the single dish fluxes. 
We found that including this 4th component with these characteristics does not improve the fits. A component with a large offset (more than 100 mas) and with similar or smaller size than our component 3 would produce phase steps in the 8 - 13 $\mu$m region that are not observed in the data. 
In fact the existence of this component is consistent with the short baseline data only
if it is so large as to be essentially resolved out by the interferometer. 
This places lower limits of $\sim 160$ mas and $\sim 200$ mas for the minor and major axis respectively.

%
\subsection{Cross-identification of the components}
\label{subsec:identification}

Previous single dish observations \citep{2000AJ....120.2904B} have  clearly revealed an elongated region of the mid infrared emission which extends up to 1" to the north but their resolution was not sufficient to resolve the central emission of the core. From our interferometric observations we  inferred that the emission of the core can be divided into two distinct regions: 
one consistent of a hot emission surrounded by warm dust (1st and 2nd components) and 
a large warm diffuse region approximately 100 mas ($\sim$ 7 pc) away from the other. 
We do not have absolute astrometric information about these components and cannot identify one with the nucleus without further assumptions.

We in fact identify the first, most compact component with the nucleus for the following reasons: the hot temperature obtained for the small component is consistent with the temperatures obtained from dust in  thermal equilibrium  close to the sublimation radius in dusty clump tori \citep{2008A&A...482...67S} and also consistent with temperature of the warm  (>600K) molecular dust that is expected to embedded the water masers \citep{1994ApJ...436L.127N}. If we attempt to locate the nucleus in components
2 or 3 it is difficult to explain the high temperature of component 1 without invoking a new, powerful heating source.
Additionally the size and orientation of component 1 closely match those of the water maser, which must be orbiting the nucleus.  

With this identification we note that at lower resolution, the peak of the mid-infrared emission will not be at the position of the small hot dust component, but near to our 3rd component, whose flux exceeds that of the nucleus by a factor of 3.5 at 12$\mu$m. 
In Fig.~\ref{sketch} we sketch the positions of our components on the contour map of \citep{2000AJ....120.2904B}. We observe that the inclination of our large component is roughly similar to the bend that is located to the north of the 12 $\mu$m image.  Our components 1/2/3 cannot be resolved in
this image.  \citet{2006A&A...446..813G} show deconvolved adaptive optics images of this
region taken with the NAOS-CONICA instrument at K-, L- and M-bands.  It is possible that our
component 3 is identical with the component labelled IR-CN on their Figure 4a, but the
limits of the resolution and dynamic range of this data make this identification uncertain.

Our estimates of the total flux for the large component show a similar behaviour as  the results of \citet{2007A&A...472..823P} who found that part of  a large north-south
component enters MIDI field of view and   contributes with 83 \% of the emission of the 
MIR from the core (within 500 mas).  The ratios of the 12 $\mu$m emission for our components can be summarized approximately as 1 : 2 : 3 for the 1st, 2nd and 3rd components respectively, meaning that the flux due to the large component should be  55 \% of the core emission. The differences in percentage is possibly due to the additional components outside
the central MIDI field of view that contribute to the component found by these
authors.

This is not the first time that a large MIR emission along or close to the polar
region is observed in AGNs. Recently interferometric observations of Circinus
\citep{2012JPhCS.372a2035T}, NGC 424 \citep{2012ApJ...755..149H} and NGC 3783
\citep{2013ApJ...771...87H} indicate the presence of mid-IR emission along the
outflow direction  with a fairly big contribution to the total flux. In
\citet{2012ApJ...755..149H} a radiatively-driven dusty wind scenario was proposed to
explain the large contributions to the mid-IR flux by polar dust.

\section{The energetics of the mid-infrared emission}
\label{sec:energetics}
 
The primary scientific results from these observations are the detection
of the ``intermediate" components 2 and 3 --1.3 and 7  parsecs north
of the core--, and the non-detection of the Tongue, about 35 parsec to the north.
In this section we consider the possible heating mechanisms for the dust in 
these components.

The usual suspects are radiative heating and shock heating.  In fact these mechanisms
collaborate.  The hot gas in a strong shock will be destroy the local dust by
sputtering and conductive heating, but it will also emit ultraviolet 
light that efficiently heats dust in the surrounding environment.
The morphology of the emission from the Tongue region supports this combined scenario.
The VLBA radio images, 
\citep{2004ApJ...613..794G}, 
show a small bright component (C) with a sharp edge near this position, suggesting a shock. 
Most of the radio emission comes from a region less than 30 mas in diameter.  
Our data, and the images from \cite{2006A&A...446..813G} indicate that the
dust emission is coming from a much larger region, probably displaced
from radio component C.  In particular the MIDI
data excludes a narrow ridge morphology that might be associated with a shock.
This extended emission presumably arises from radiatively heated dust.

\cite{2012ApJ...756..180W} describe a similar scenario based on relatively high resolution 
(300 mas) Chandra X-ray data.  The X-ray and radio bright region HST-G about 1"
north of the nucleus shows an X-ray spectrum 
containing both photoionized and high-density thermal components.  Detailed X-ray spectra for
the other X-ray components in the region are not available, but the ratio of [OIII] to soft
X-ray continuum indicates that some (labelled HST-D, E, F) are radiation heated, while
others (HST-G, H and the near-nuclear region HST-A, B, C) contain shocked gas.
The HST-A, B, C region contains the nucleus (to the extent not blocked by Compton scattering),
our components 1,2, and 3, and the Tongue.  Unfortunately the spatial resolution of
the X-ray and [OIII] data cannot distinguish between these subcomponents.  The very high
resolution VLBA data of \cite{2004ApJ...613..794G} show a flat-spectrum nuclear component,
presumably coinciding with out component 1, but no emission at our positions 2 or 3.
They find  strong synchrotron emission at the Tongue and at
their NE component, which curiously shows no enhanced X-ray, [OIII] or infrared emission.
There are several regions, e.g. HST-D, E, F of \cite{1991ApJ...369L..27E} 
that show X-ray, [OIII] and infrared emission
but where there is no sign of shock enhancement of the synchrotron jet \citep{2004ApJ...613..794G}.
Regions NE-5, 6, 7 of \cite{2005MNRAS.363L...1G} show the same features.  There is no evidence
at these positions of direct interaction with the radio jet, although they lie at the
edge of a radio cocoon \citep{1983ApJ...275....8W}.

This summary indicates the complexity of the region and suggests that different
mechanisms dominate at different positions.  The data from the Tongue region seems
to support the shock plus radiative heating in this area.  On the
other hand, our region 3 show no signature of shocks in the radio. This fact and the
proximity to the nucleus favor heating by UV-radiation from the nucleus.

We can examine whether the infrared spectral information in the region is consistent
with this hypothesis.
The luminosity produced by the nucleus is sufficient to obtain the dust temperature measured
at this position. The expected temperature of dust (assuming silicate grains) 
heated directly by the central engine is given by:

\begin{equation}
 T\simeq 1500 \left(\frac{L_{uv,46}}{r_{pc}^2}\right)^\frac{1}{5.6} K
\end{equation}

where $L_{uv,46}$ is the luminosity of the heating source in units of $10^{46}$ erg s$^{-1}$ \citep{1987ApJ...320..537B}. For the central source of  NGC 1068 we take the UV luminosity $L_{uv}=1.5\times 10^{45}$ erg s$^{-1}$ previously used by \citet{2003A&A...411..335G} to reproduce the central K band flux and continuum. Dust at a distance of $r\sim 7$ pc (100 mas) 
can be heated to $T=530$ K. The color temperatures in our wavelength range are lower, $\sim 400$ K.

The spectra of the various infrared components show various, sometime quite high
color temperatures, but it is difficult to use this to distinguish radiative from
shock heating.  The dust in the shock heated Tongue region shows short wavelength fluxes with
color temperatures $\sim 700$ K \citep{2006A&A...446..813G}, but some of the shortest wavelength
data may represent scattered nuclear light rather than local thermal emission.
The spatial resolution of the data in \cite{2006A&A...446..813G} 
is not sufficient to unambiguously determine a short wavelength
color temperature for our component 3.  They report aperture
fluxes for the non-deconvolved K-band image of the nucleus that increase 
from $\sim 70$ mJy at 80 mas radius to $\sim 130$ mJy at 130 mas 
radius and $\sim 190$ mJy at 270 mas.  If we extrapolate our
N-band flux of component 3 to the K-band with the same fluxes ratio as the Tongue region,
it would have a flux of $\sim 50$ mJy, which is consistent with the fluxes just quoted.
The speckle data of \cite{2004A&A...425...77W} also give
some indication of short wavelength radiation from our component but the resolution is
again marginal, and the same problem of scattered light exists.
In conclusion the infrared data do not exclude N- to K-band color temperatures up to $\sim 700$ K,
which would be difficult to explain by radiative heating from the nucleus.

We anticipate that high-resolution interferometric mapping of NGC~1068 
at 3$\mu$m and 5$\mu$m with the MATISSE interferometer \citep{2008SPIE.7013E..70L},
will determine accurate infrared spectra for these regions and allow 
disentanglement of the thermal emission and scattering.  Here we assume
on the basis of the lack of radio emission in component 3 that it is
heated radiatively by the nucleus.  
This component emits about 60\% of
the near nuclear mid-infrared flux. Hence this structure must absorb most
of the nuclear UV-emission.  Thus the dust that extinguishes the nuclear
emission in this Seyfert 2 galaxy is not distributed in a disk- or torus-like
structure, cloudy or otherwise, but in a narrow cylindrical or conical structure
generally in the jet direction, which is, however, not symmetric around the jet. 
\cite{2011ApJ...739...69M} found similarly that the coronal line ionization cone
in NGC~1068 was also quite narrow.  They estimate the half-opening angle to the north
to be $\sim 27^\circ$.  If the simple unified models of Seyfert galaxies is in fact true,
and the opening angles for viewing the direct emission from the nucleus were as small as
that in NGC~1068, only about 10\% of Seyfert would be of Type 1, instead of the observed
value of $\sim 50$\%. A spread in the values for the opening angles was earlier noted by studying the kinematics of the narrow line region using the HST \citep{2013ApJS..209....1F} and VLT/SINFONI \citep{2011ApJ...739...69M} and for the case of NGC~1068 the relative narrow opening angle of the ionization cone reported in both studies are in agreement with our observations. Thus there is evidence for a large spread of the opening angles of AGNs, and that the actual difference between Types 1 and 2 may be more complicated than a simple inclination angle effect.

\section{The north/south asymmetry}
\label{sec:asymmetry}

It is noteworthy that the majority of the emission in the immediate circumnuclear neighborhood is highly asymmetric, lying to the north of the nucleus itself.  This is the same side where most of the radio, near-infrared emission and optical ''ionization cone'' emission is seen. At shorter wavelengths, the asymmetry is usually attributed to foreground absorption from the inclined dust structure, but a similar mechanism in the mid-infrared requires very high dust opacities.  Suppose that an additional component identical to component 3 were present in the south.  
We would need several optical depths at 12$\mu$m at a position 7 parsec south of the nucleus to obscure this component.  This dust needs to be quite cold, e.g. $<$ 200 K, 
in order that its own emission not dominate our picture.  A similar picture, but with
lower forground opacity is presented by \cite{2013arXiv1312.4534T} for Circinus.  
In view of the complexity of this source, it cannot be ruled out that
a cold cloud or disk structure at large radius extinguishes the southern infrared emission.  
This would not explain the north/south asymmetry of the radio emission, 
this may represent accidental circumstances of
the interception of the jet by clouds. 

\cite{2013arXiv1312.4534T} have found a similar
picture in the Circinus galaxy.  They find non-torus emission aligned with
radio jet.  The mid-infrared continuum emission is more or less symmetric in these regions,
but the $10\mu$m silicate absorption is much stronger on the side where optical emission
is missing.  This may represent a case where the foreground dust thickness is enough to
block the center of the silicate band, but not enough to absorb most of the N-band continuum.
In this picture the dust column density in NGC 1068 would be at least three times higher, so
that both the continuum and the silicate feature are blocked.

\section{Conclusions}
\label{sec:conclusions}

We present new interferometric data from MIDI/VLTI using the Auxiliary Telescopes that
allow us to investigate the structure of  the 5 - 10 pc scale mid-infrared nuclear emission from NGC 1068. Our observations complement the previous UT interferometric observations, that trace the small scales structures,  and the single dish images of the extended region. The ATs, which have an effective area smaller by a factor of $\sim 20$ than the UTs,
were able to track fringes for NGC 1068 measuring fluxes close to the limit of the instrument.

We summarize our results as follows:

\begin{itemize}
\item Fits using non-concentric gaussian components indicate that most of the mid-infrared emission is attributed to the large scale structures. This emission is associated with warm dust distributed in two major components, one close to
the center  and one with a distance  larger than 80 mas and close to  16$^\circ$ -
18$^\circ$ in the NW direction. The central warm region would mostly be seen as the
extension of the hot emission region, while the offset region may be attributed to
dusty clouds located close the northern ionization cone. 
The parameters found for this component indicate a size of $\sim 14$ parsec, strong elongation along a position angle of $\sim -35^\circ$ and a contribution to the total flux at 12 $\mu$m more than  three times that of the central hot region.
 \item The non-zero chromatic phases on the shorter baselines indicate major asymmetries in the emission on the 10 parsec scale, primarily along the north-south axis.
\item The 3-gaussian model does not fit well the intermediate baseline data, probably due to the complexity of the structures in the equivalent size scales.
This, combined with the limited information from the chromatic phases,
leads to uncertainty in description of the structures on equivalent scales. The displacement, size, and orientation of the large northern component is determined        from the data at the shortest baselines, where the quality of the gaussian fits is good.  
\item The ''low resolution'' picture derived from the gaussian fits indicates that the near-nuclear ($< 10$ pc) emission is strongly asymmetric with respect to the nucleus, with the major components to the north agreeing in orientation with those seen in near-infrared NACO images.  The north-south asymmetric may be caused by strong extinction of the southern components, or by intrinsic asymmetry
of the dust components.
\item The warm component 3, $\sim 7$ parsecs north of the hotter nuclear disk, apparently intercepts a large fraction of the nuclear UV-emission.  Thus there are several obscuring components at different disk ''latitudes'' that can cause Seyfert 2 appearance in AGNs.  The volume that is heated by this
emission is quite narrow; the viewing angles from which this  galaxy would be classified as Seyfert 1 cover only $\sim 10$\% of the sky.
 \item We  do not observe evidence for variability effects in the mid-infrared emission of the small scales structures in the nuclear region of NGC 1068. Similar $(u,v)$ points observed with a difference of 7 years do not show significant differences (less than 10 \%).
 \item Independently calibrated observations with the ATs over periods of several days show a high level of repeatability.
\end{itemize}

\begin{acknowledgements}
The authors thank the anonymous referee for the thoughtful and helpful comments. 
This work was based on observations with ESO telescopes at the La Silla Paranal
Observatory under programs: 074.A-9015, 076.B-0743, 277.B-5014, 080.B-0928, 087.C-0824 and 089.B-0099.
N. L\'opez-Gonzaga was supported by grant 614.000 from the Nederlandse Organisatie voor
Wetenschappelijk Onderzoek and acknowledges support from a CONACyT graduate fellowship.
\end{acknowledgements}

\bibliographystyle{aa} 
\bibliography{paper1}

\begin{thebibliography}{63}
\expandafter\ifx\csname natexlab\endcsname\relax\def\natexlab#1{#1}\fi

\bibitem[{{Antonucci}(1993)}]{1993ARA&A..31..473A}
{Antonucci}, R. 1993, \araa, 31, 473

\bibitem[{{Antonucci}(1984)}]{1984ApJ...278..499A}
{Antonucci}, R.~R.~J. 1984, \apj, 278, 499

\bibitem[{{Antonucci} \& {Miller}(1985)}]{1985ApJ...297..621A}
{Antonucci}, R.~R.~J. \& {Miller}, J.~S. 1985, \apj, 297, 621

\bibitem[{{Barvainis}(1987)}]{1987ApJ...320..537B}
{Barvainis}, R. 1987, \apj, 320, 537

\bibitem[{{Beckert} {et~al.}(2008){Beckert}, {Driebe}, {H{\"o}nig}, \&
  {Weigelt}}]{2008A&A...486L..17B}
{Beckert}, T., {Driebe}, T., {H{\"o}nig}, S.~F., \& {Weigelt}, G. 2008, \aap,
  486, L17

\bibitem[{{Bock} {et~al.}(1998){Bock}, {Marsh}, {Ressler}, \&
  {Werner}}]{1998ApJ...504L...5B}
{Bock}, J.~J., {Marsh}, K.~A., {Ressler}, M.~E., \& {Werner}, M.~W. 1998,
  \apjl, 504, L5

\bibitem[{{Bock} {et~al.}(2000){Bock}, {Neugebauer}, {Matthews}, {Soifer},
  {Becklin}, {Ressler}, {Marsh}, {Werner}, {Egami}, \&
  {Blandford}}]{2000AJ....120.2904B}
{Bock}, J.~J., {Neugebauer}, G., {Matthews}, K., {et~al.} 2000, \aj, 120, 2904

\bibitem[{{Burtscher} {et~al.}(2009){Burtscher}, {Jaffe}, {Raban},
  {Meisenheimer}, {Tristram}, \& {R{\"o}ttgering}}]{2009ApJ...705L..53B}
{Burtscher}, L., {Jaffe}, W., {Raban}, D., {et~al.} 2009, \apjl, 705, L53

\bibitem[{{Burtscher} {et~al.}(2010){Burtscher}, {Meisenheimer}, {Jaffe},
  {Tristram}, \& {R{\"o}ttgering}}]{2010PASA...27..490B}
{Burtscher}, L., {Meisenheimer}, K., {Jaffe}, W., {Tristram}, K.~R.~W., \&
  {R{\"o}ttgering}, H.~J.~A. 2010, \pasa, 27, 490

\bibitem[{{Burtscher} {et~al.}(2013){Burtscher}, {Meisenheimer}, {Tristram},
  {Jaffe}, {H{\"o}nig}, {Davies}, {Kishimoto}, {Pott}, {R{\"o}ttgering},
  {Schartmann}, {Weigelt}, \& {Wolf}}]{2013A&A...558A.149B}
{Burtscher}, L., {Meisenheimer}, K., {Tristram}, K.~R.~W., {et~al.} 2013, \aap,
  558, A149

\bibitem[{{Burtscher} {et~al.}(2012){Burtscher}, {Tristram}, {Jaffe}, \&
  {Meisenheimer}}]{2012SPIE.8445E..1GB}
{Burtscher}, L., {Tristram}, K.~R.~W., {Jaffe}, W.~J., \& {Meisenheimer}, K.
  2012, in Society of Photo-Optical Instrumentation Engineers (SPIE) Conference
  Series, Vol. 8445, Society of Photo-Optical Instrumentation Engineers (SPIE)
  Conference Series

\bibitem[{{Cohen} {et~al.}(1999){Cohen}, {Walker}, {Carter}, {Hammersley},
  {Kidger}, \& {Noguchi}}]{1999AJ....117.1864C}
{Cohen}, M., {Walker}, R.~G., {Carter}, B., {et~al.} 1999, \aj, 117, 1864

\bibitem[{{Deroo} {et~al.}(2007){Deroo}, {van Winckel}, {Verhoelst}, {Min},
  {Reyniers}, \& {Waters}}]{2007A&A...467.1093D}
{Deroo}, P., {van Winckel}, H., {Verhoelst}, T., {et~al.} 2007, \aap, 467, 1093

\bibitem[{{Evans} {et~al.}(1991){Evans}, {Ford}, {Kinney}, {Antonucci},
  {Armus}, \& {Caganoff}}]{1991ApJ...369L..27E}
{Evans}, I.~N., {Ford}, H.~C., {Kinney}, A.~L., {et~al.} 1991, \apjl, 369, L27

\bibitem[{{Fischer} {et~al.}(2013){Fischer}, {Crenshaw}, {Kraemer}, \&
  {Schmitt}}]{2013ApJS..209....1F}
{Fischer}, T.~C., {Crenshaw}, D.~M., {Kraemer}, S.~B., \& {Schmitt}, H.~R.
  2013, \apjs, 209, 1

\bibitem[{{Galliano} {et~al.}(2005){Galliano}, {Pantin}, {Alloin}, \&
  {Lagage}}]{2005MNRAS.363L...1G}
{Galliano}, E., {Pantin}, E., {Alloin}, D., \& {Lagage}, P.~O. 2005, \mnras,
  363, L1

\bibitem[{{Gallimore} {et~al.}(1996){Gallimore}, {Baum}, \&
  {O'Dea}}]{1996ApJ...464..198G}
{Gallimore}, J.~F., {Baum}, S.~A., \& {O'Dea}, C.~P. 1996, \apj, 464, 198

\bibitem[{{Gallimore} {et~al.}(2004){Gallimore}, {Baum}, \&
  {O'Dea}}]{2004ApJ...613..794G}
{Gallimore}, J.~F., {Baum}, S.~A., \& {O'Dea}, C.~P. 2004, \apj, 613, 794

\bibitem[{{Glass}(1997)}]{1997Ap&SS.248..191G}
{Glass}, I.~S. 1997, \apss, 248, 191

\bibitem[{{Gratadour} {et~al.}(2003){Gratadour}, {Cl{\'e}net}, {Rouan}, {Lai},
  \& {Forveille}}]{2003A&A...411..335G}
{Gratadour}, D., {Cl{\'e}net}, Y., {Rouan}, D., {Lai}, O., \& {Forveille}, T.
  2003, \aap, 411, 335

\bibitem[{{Gratadour} {et~al.}(2006){Gratadour}, {Rouan}, {Mugnier}, {Fusco},
  {Cl{\'e}net}, {Gendron}, \& {Lacombe}}]{2006A&A...446..813G}
{Gratadour}, D., {Rouan}, D., {Mugnier}, L.~M., {et~al.} 2006, \aap, 446, 813

\bibitem[{{Greenhill} {et~al.}(1996){Greenhill}, {Gwinn}, {Antonucci}, \&
  {Barvainis}}]{1996ApJ...472L..21G}
{Greenhill}, L.~J., {Gwinn}, C.~R., {Antonucci}, R., \& {Barvainis}, R. 1996,
  \apjl, 472, L21

\bibitem[{{H{\"o}nig} {et~al.}(2012){H{\"o}nig}, {Kishimoto}, {Antonucci},
  {Marconi}, {Prieto}, {Tristram}, \& {Weigelt}}]{2012ApJ...755..149H}
{H{\"o}nig}, S.~F., {Kishimoto}, M., {Antonucci}, R., {et~al.} 2012, \apj, 755,
  149

\bibitem[{{H{\"o}nig} {et~al.}(2013){H{\"o}nig}, {Kishimoto}, {Tristram},
  {Prieto}, {Gandhi}, {Asmus}, {Antonucci}, {Burtscher}, {Duschl}, \&
  {Weigelt}}]{2013ApJ...771...87H}
{H{\"o}nig}, S.~F., {Kishimoto}, M., {Tristram}, K.~R.~W., {et~al.} 2013, \apj,
  771, 87

\bibitem[{{Jaffe} {et~al.}(2004){Jaffe}, {Meisenheimer}, {R{\"o}ttgering},
  {Leinert}, {Richichi}, {Chesneau}, {Fraix-Burnet}, {Glazenborg-Kluttig},
  {Granato}, {Graser}, {Heijligers}, {K{\"o}hler}, {Malbet}, {Miley},
  {Paresce}, {Pel}, {Perrin}, {Przygodda}, {Schoeller}, {Sol}, {Waters},
  {Weigelt}, {Woillez}, \& {de Zeeuw}}]{2004Natur.429...47J}
{Jaffe}, W., {Meisenheimer}, K., {R{\"o}ttgering}, H.~J.~A., {et~al.} 2004,
  \nat, 429, 47

\bibitem[{{Jaffe}(2004)}]{2004SPIE.5491..715J}
{Jaffe}, W.~J. 2004, in Society of Photo-Optical Instrumentation Engineers
  (SPIE) Conference Series, Vol. 5491, Society of Photo-Optical Instrumentation
  Engineers (SPIE) Conference Series, ed. W.~A. {Traub}, 715

\bibitem[{{Kemper} {et~al.}(2004){Kemper}, {Vriend}, \&
  {Tielens}}]{2004ApJ...609..826K}
{Kemper}, F., {Vriend}, W.~J., \& {Tielens}, A.~G.~G.~M. 2004, \apj, 609, 826

\bibitem[{{Kishimoto} {et~al.}(2009){Kishimoto}, {H{\"o}nig}, {Tristram}, \&
  {Weigelt}}]{2009A&A...493L..57K}
{Kishimoto}, M., {H{\"o}nig}, S.~F., {Tristram}, K.~R.~W., \& {Weigelt}, G.
  2009, \aap, 493, L57

\bibitem[{{K{\"o}hler} \& {Li}(2010)}]{2010MNRAS.406L...6K}
{K{\"o}hler}, M. \& {Li}, A. 2010, \mnras, 406, L6

\bibitem[{{Laor} \& {Draine}(1993)}]{1993ApJ...402..441L}
{Laor}, A. \& {Draine}, B.~T. 1993, \apj, 402, 441

\bibitem[{{Leinert} {et~al.}(2003){Leinert}, {Graser}, {Przygodda}, {Waters},
  {Perrin}, {Jaffe}, {Lopez}, {Bakker}, {B{\"o}hm}, {Chesneau}, {Cotton},
  {Damstra}, {de Jong}, {Glazenborg-Kluttig}, {Grimm}, {Hanenburg}, {Laun},
  {Lenzen}, {Ligori}, {Mathar}, {Meisner}, {Morel}, {Morr}, {Neumann}, {Pel},
  {Schuller}, {Rohloff}, {Stecklum}, {Storz}, {von der L{\"u}he}, \&
  {Wagner}}]{2003Ap&SS.286...73L}
{Leinert}, C., {Graser}, U., {Przygodda}, F., {et~al.} 2003, \apss, 286, 73

\bibitem[{{Lopez} {et~al.}(2008){Lopez}, {Antonelli}, {Wolf}, {Lagarde},
  {Jaffe}, {Navarro}, {Graser}, {Petrov}, {Weigelt}, {Bresson}, {Hofmann},
  {Beckman}, {Henning}, {Laun}, {Leinert}, {Kraus}, {Robbe-Dubois}, {Vakili},
  {Richichi}, {Abraham}, {Augereau}, {Behrend}, {Berio}, {Berruyer},
  {Chesneau}, {Clausse}, {Connot}, {Demyk}, {Danchi}, {Dugu{\'e}}, {Finger},
  {Flament}, {Glazenborg}, {Hannenburg}, {Heininger}, {Hugues}, {Hron},
  {Jankov}, {Kerschbaum}, {Kroes}, {Linz}, {Lizon}, {Mathias}, {Mathar},
  {Matter}, {Menut}, {Meisenheimer}, {Millour}, {Nardetto}, {Neumann},
  {Nussbaum}, {Niedzielski}, {Mosoni}, {Olofsson}, {Rabbia}, {Ratzka}, {Rigal},
  {Roussel}, {Schertl}, {Schmider}, {Stecklum}, {Thiebaut}, {Vannier}, {Valat},
  {Wagner}, \& {Waters}}]{2008SPIE.7013E..70L}
{Lopez}, B., {Antonelli}, P., {Wolf}, S., {et~al.} 2008, in Society of
  Photo-Optical Instrumentation Engineers (SPIE) Conference Series, Vol. 7013,
  Society of Photo-Optical Instrumentation Engineers (SPIE) Conference Series

\bibitem[{{Meisenheimer} {et~al.}(2007){Meisenheimer}, {Tristram}, {Jaffe},
  {Israel}, {Neumayer}, {Raban}, {R{\"o}ttgering}, {Cotton}, {Graser},
  {Henning}, {Leinert}, {Lopez}, {Perrin}, \& {Prieto}}]{2007A&A...471..453M}
{Meisenheimer}, K., {Tristram}, K.~R.~W., {Jaffe}, W., {et~al.} 2007, \aap,
  471, 453

\bibitem[{{M{\"u}ller} {et~al.}(2010){M{\"u}ller}, {Pott}, {Morel}, {Abuter},
  {van Belle}, {van Boekel}, {Burtscher}, {Delplancke}, {Henning}, {Jaffe},
  {Leinert}, {Lopez}, {Matter}, {Meisenheimer}, {Schmid}, {Tristram}, \&
  {Verhoeff}}]{2010SPIE.7734E..60M}
{M{\"u}ller}, A., {Pott}, J.-U., {Morel}, S., {et~al.} 2010, in Society of
  Photo-Optical Instrumentation Engineers (SPIE) Conference Series, Vol. 7734,
  Society of Photo-Optical Instrumentation Engineers (SPIE) Conference Series

\bibitem[{{M{\"u}ller S{\'a}nchez} {et~al.}(2009){M{\"u}ller S{\'a}nchez},
  {Davies}, {Genzel}, {Tacconi}, {Eisenhauer}, {Hicks}, {Friedrich}, \&
  {Sternberg}}]{2009ApJ...691..749M}
{M{\"u}ller S{\'a}nchez}, F., {Davies}, R.~I., {Genzel}, R., {et~al.} 2009,
  \apj, 691, 749

\bibitem[{{M{\"u}ller-S{\'a}nchez} {et~al.}(2011){M{\"u}ller-S{\'a}nchez},
  {Prieto}, {Hicks}, {Vives-Arias}, {Davies}, {Malkan}, {Tacconi}, \&
  {Genzel}}]{2011ApJ...739...69M}
{M{\"u}ller-S{\'a}nchez}, F., {Prieto}, M.~A., {Hicks}, E.~K.~S., {et~al.}
  2011, \apj, 739, 69

\bibitem[{{Mutschke} {et~al.}(1998){Mutschke}, {Begemann}, {Dorschner},
  {Guertler}, {Gustafson}, {Henning}, \& {Stognienko}}]{1998A&A...333..188M}
{Mutschke}, H., {Begemann}, B., {Dorschner}, J., {et~al.} 1998, \aap, 333, 188

\bibitem[{{Nenkova} {et~al.}(2008){Nenkova}, {Sirocky}, {Ivezi{\'c}}, \&
  {Elitzur}}]{2008ApJ...685..147N}
{Nenkova}, M., {Sirocky}, M.~M., {Ivezi{\'c}}, {\v Z}., \& {Elitzur}, M. 2008,
  \apj, 685, 147

\bibitem[{{Neufeld} {et~al.}(1994){Neufeld}, {Maloney}, \&
  {Conger}}]{1994ApJ...436L.127N}
{Neufeld}, D.~A., {Maloney}, P.~R., \& {Conger}, S. 1994, \apjl, 436, L127

\bibitem[{{Poncelet} {et~al.}(2007){Poncelet}, {Doucet}, {Perrin}, {Sol}, \&
  {Lagage}}]{2007A&A...472..823P}
{Poncelet}, A., {Doucet}, C., {Perrin}, G., {Sol}, H., \& {Lagage}, P.~O. 2007,
  \aap, 472, 823

\bibitem[{{Poncelet} {et~al.}(2006){Poncelet}, {Perrin}, \&
  {Sol}}]{2006A&A...450..483P}
{Poncelet}, A., {Perrin}, G., \& {Sol}, H. 2006, \aap, 450, 483

\bibitem[{{Poncelet} {et~al.}(2008){Poncelet}, {Sol}, \&
  {Perrin}}]{2008A&A...481..305P}
{Poncelet}, A., {Sol}, H., \& {Perrin}, G. 2008, \aap, 481, 305

\bibitem[{{Pott} {et~al.}(2012){Pott}, {M{\"u}ller}, {Karovicova}, \&
  {Delplancke}}]{2012SPIE.8445E..0QP}
{Pott}, J.-U., {M{\"u}ller}, A., {Karovicova}, I., \& {Delplancke}, F. 2012, in
  Society of Photo-Optical Instrumentation Engineers (SPIE) Conference Series,
  Vol. 8445, Society of Photo-Optical Instrumentation Engineers (SPIE)
  Conference Series

\bibitem[{{Raban} {et~al.}(2009){Raban}, {Jaffe}, {R{\"o}ttgering},
  {Meisenheimer}, \& {Tristram}}]{2009MNRAS.394.1325R}
{Raban}, D., {Jaffe}, W., {R{\"o}ttgering}, H., {Meisenheimer}, K., \&
  {Tristram}, K.~R.~W. 2009, \mnras, 394, 1325

\bibitem[{{Rouan} {et~al.}(2004){Rouan}, {Lacombe}, {Gendron}, {Gratadour},
  {Cl{\'e}net}, {Lagrange}, {Mouillet}, {Boisson}, {Rousset}, {Fusco},
  {Mugnier}, {S{\'e}chaud}, {Thatte}, {Genzel}, {Gigan}, {Arsenault}, \&
  {Kern}}]{2004A&A...417L...1R}
{Rouan}, D., {Lacombe}, F., {Gendron}, E., {et~al.} 2004, \aap, 417, L1

\bibitem[{{Rouan} {et~al.}(1998){Rouan}, {Rigaut}, {Alloin}, {Doyon}, {Lai},
  {Crampton}, {Gendron}, \& {Arsenault}}]{1998A&A...339..687R}
{Rouan}, D., {Rigaut}, F., {Alloin}, D., {et~al.} 1998, \aap, 339, 687

\bibitem[{{Sanders} {et~al.}(1989){Sanders}, {Phinney}, {Neugebauer}, {Soifer},
  \& {Matthews}}]{1989ApJ...347...29S}
{Sanders}, D.~B., {Phinney}, E.~S., {Neugebauer}, G., {Soifer}, B.~T., \&
  {Matthews}, K. 1989, \apj, 347, 29

\bibitem[{{Schartmann} {et~al.}(2010){Schartmann}, {Burkert}, {Krause},
  {Camenzind}, {Meisenheimer}, \& {Davies}}]{2010MNRAS.403.1801S}
{Schartmann}, M., {Burkert}, A., {Krause}, M., {et~al.} 2010, \mnras, 403, 1801

\bibitem[{{Schartmann} {et~al.}(2008){Schartmann}, {Meisenheimer}, {Camenzind},
  {Wolf}, {Tristram}, \& {Henning}}]{2008A&A...482...67S}
{Schartmann}, M., {Meisenheimer}, K., {Camenzind}, M., {et~al.} 2008, \aap,
  482, 67

\bibitem[{{Schartmann} {et~al.}(2009){Schartmann}, {Meisenheimer}, {Klahr},
  {Camenzind}, {Wolf}, \& {Henning}}]{2009MNRAS.393..759S}
{Schartmann}, M., {Meisenheimer}, K., {Klahr}, H., {et~al.} 2009, \mnras, 393,
  759

\bibitem[{{Seyfert}(1943)}]{1943ApJ....97...28S}
{Seyfert}, C.~K. 1943, \apj, 97, 28

\bibitem[{{Taranova} \& {Shenavrin}(2006)}]{2006A&AT...25..233T}
{Taranova}, O.~G. \& {Shenavrin}, V.~I. 2006, Astronomical and Astrophysical
  Transactions, 25, 233

\bibitem[{{Tomono} {et~al.}(2001){Tomono}, {Doi}, {Usuda}, \&
  {Nishimura}}]{2001ApJ...557..637T}
{Tomono}, D., {Doi}, Y., {Usuda}, T., \& {Nishimura}, T. 2001, \apj, 557, 637

\bibitem[{{Tristram} {et~al.}(2013){Tristram}, {Burtscher}, {Jaffe},
  {Meisenheimer}, {H{\"o}nig}, {Kishimoto}, {Schartmann}, \&
  {Weigelt}}]{2013arXiv1312.4534T}
{Tristram}, K.~R.~W., {Burtscher}, L., {Jaffe}, W., {et~al.} 2013, ArXiv
  e-prints

\bibitem[{{Tristram} {et~al.}(2007){Tristram}, {Meisenheimer}, {Jaffe},
  {Schartmann}, {Rix}, {Leinert}, {Morel}, {Wittkowski}, {R{\"o}ttgering},
  {Perrin}, {Lopez}, {Raban}, {Cotton}, {Graser}, {Paresce}, \&
  {Henning}}]{2007A&A...474..837T}
{Tristram}, K.~R.~W., {Meisenheimer}, K., {Jaffe}, W., {et~al.} 2007, \aap,
  474, 837

\bibitem[{{Tristram} {et~al.}(2009){Tristram}, {Raban}, {Meisenheimer},
  {Jaffe}, {R{\"o}ttgering}, {Burtscher}, {Cotton}, {Graser}, {Henning},
  {Leinert}, {Lopez}, {Morel}, {Perrin}, \& {Wittkowski}}]{2009A&A...502...67T}
{Tristram}, K.~R.~W., {Raban}, D., {Meisenheimer}, K., {et~al.} 2009, \aap,
  502, 67

\bibitem[{{Tristram} {et~al.}(2012){Tristram}, {Schartmann}, {Burtscher},
  {Meisenheimer}, {Jaffe}, {Kishimoto}, {H{\"o}nig}, \&
  {Weigelt}}]{2012JPhCS.372a2035T}
{Tristram}, K.~R.~W., {Schartmann}, M., {Burtscher}, L., {et~al.} 2012, Journal
  of Physics Conference Series, 372, 012035

\bibitem[{{Urry} \& {Padovani}(1995)}]{1995PASP..107..803U}
{Urry}, C.~M. \& {Padovani}, P. 1995, \pasp, 107, 803

\bibitem[{{Vollmer} {et~al.}(2008){Vollmer}, {Beckert}, \&
  {Davies}}]{2008A&A...491..441V}
{Vollmer}, B., {Beckert}, T., \& {Davies}, R.~I. 2008, \aap, 491, 441

\bibitem[{{Wang} {et~al.}(2012){Wang}, {Fabbiano}, {Karovska}, {Elvis}, \&
  {Risaliti}}]{2012ApJ...756..180W}
{Wang}, J., {Fabbiano}, G., {Karovska}, M., {Elvis}, M., \& {Risaliti}, G.
  2012, \apj, 756, 180

\bibitem[{{Weigelt} {et~al.}(2004){Weigelt}, {Wittkowski}, {Balega}, {Beckert},
  {Duschl}, {Hofmann}, {Men'shchikov}, \& {Schertl}}]{2004A&A...425...77W}
{Weigelt}, G., {Wittkowski}, M., {Balega}, Y.~Y., {et~al.} 2004, \aap, 425, 77

\bibitem[{{Wilson} \& {Ulvestad}(1983)}]{1983ApJ...275....8W}
{Wilson}, A.~S. \& {Ulvestad}, J.~S. 1983, \apj, 275, 8

\bibitem[{{Wittkowski} {et~al.}(2004){Wittkowski}, {Kervella}, {Arsenault},
  {Paresce}, {Beckert}, \& {Weigelt}}]{2004A&A...418L..39W}
{Wittkowski}, M., {Kervella}, P., {Arsenault}, R., {et~al.} 2004, \aap, 418,
  L39

\end{thebibliography}

\clearpage
\begin{appendix} 
\section{Plots of UT Data}

\begin{figure*}
  \centering
  \includegraphics[angle=0,width=0.8\textwidth]{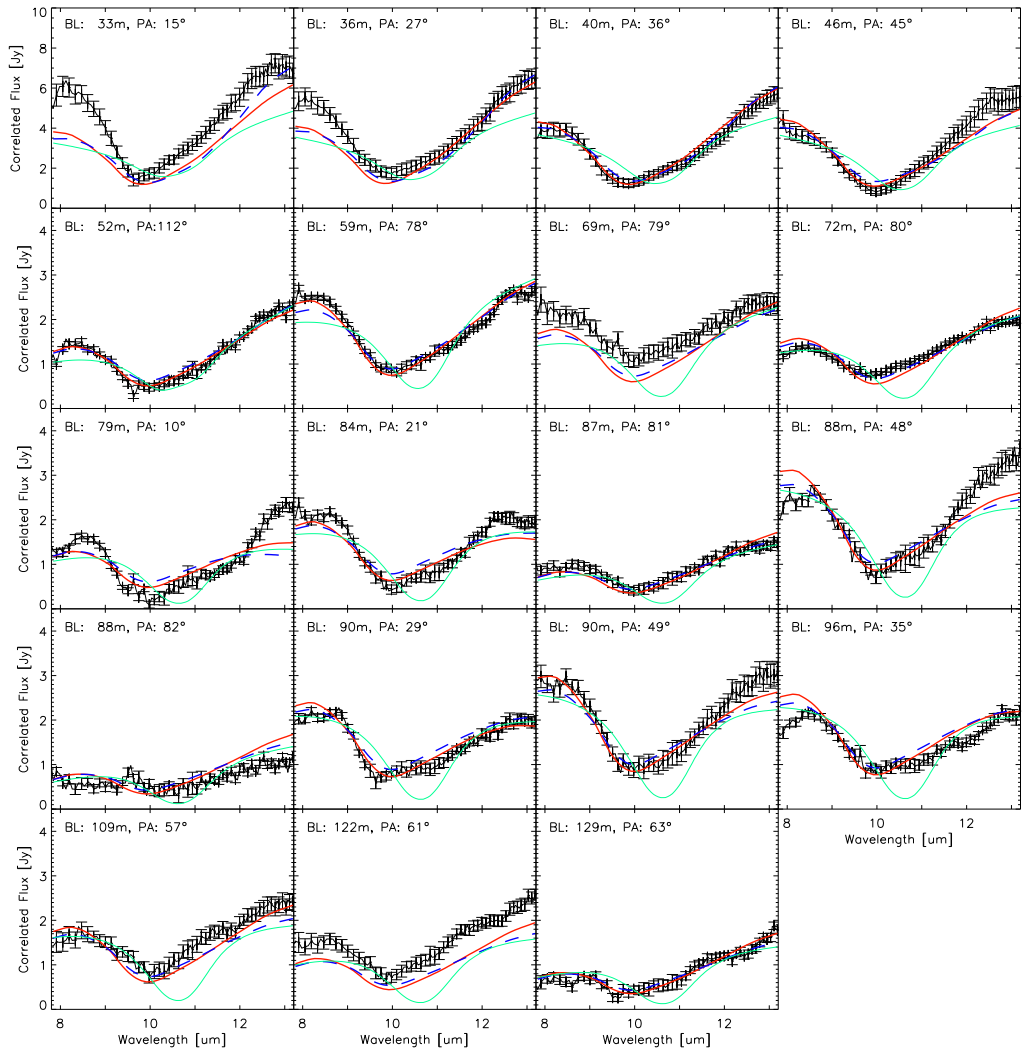}
  \caption{Correlated fluxes of observations taken with UTs. The red and dashed blue line are the curves obtained with our 1st and 2nd best fit models. The light blue curve shows the correlated fluxes of our best fit model found using SiC as the dust template for all the components. We observe that the main reason of discarding SiC as the best option is mostly due to the shape of the absorption feature.}
  \label{corrut}
\end{figure*}
\clearpage
\begin{figure*}
  \centering
  \includegraphics[angle=0,width=0.8\textwidth]{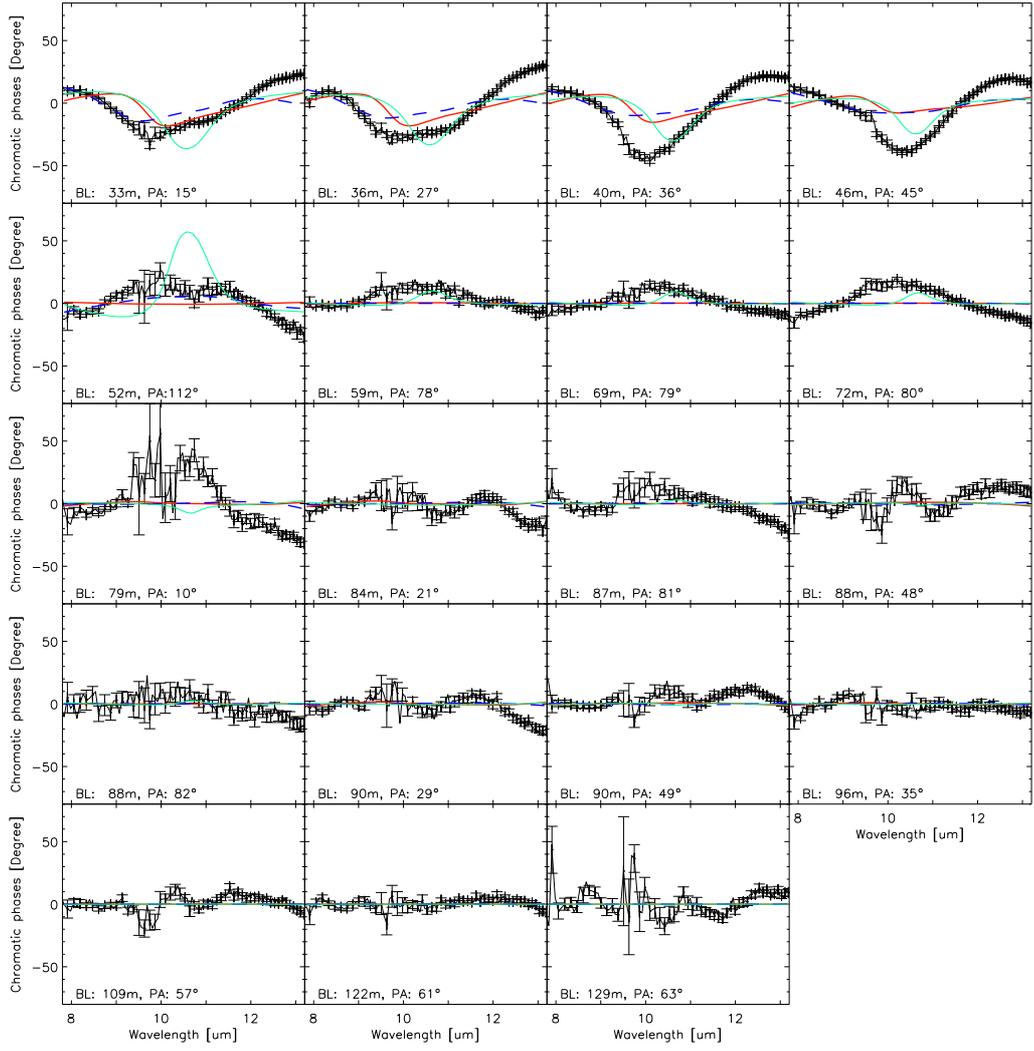}
    \caption{The same as the Fig.~\ref{corrut} but chromatic phases are plotted instead of correlated fluxes.}
    \label{phiut}
\end{figure*}
\clearpage

\section{LOG of observations}

\begin{table*}
\caption{Log of observations: NGC1068. The columns are: {\it Time} of fringe track observation, {\it BL} projected baseline length, {\it PA} position angle, name of the {\it Calibrator}, {\it Caltime} Time of the calibrator fringe track observation, {\it Airmass} of fringe track, {\it Seeing} during fringe track observation, {\it Good?} - goodness of observation (good:1, bad:0). {\it Stacked} with the following observation (yes:1, no:0) Corresponding {\it group} according to criterion of Sect.~\ref{subsec:corrfluxes} }
\centering
\begin{tabular}{c c c c c c c c c c}
\hline\hline
Time & BL [m] & PA [deg] & Calibrator & Caltime & Airmass & Seeing & Good? & Stacked & Group \T \\
\hline
\hline
\multicolumn{6}{l}{2007-10-07: E0G0} \T \B \\
06:04:14 & 15.7 &  72.4 &  HD10380 & 06:37:22 &  1.1 &  2.0 &  1& 1 & \#3\\
06:08:04 & 15.8 &  72.5 &  HD10380 & 06:37:22 &  1.1 &  2.0 &  1& 0 &\#3\\
07:06:35 & 16.0 &  72.7 &  HD10380 & 06:37:22 &  1.1 &  1.7 &  1& 1 &\#3\\
07:10:28 & 16.0 &  72.7 &  HD10380 & 06:37:22 &  1.1 &  1.9 &  1& 0 &\#3\\
08:09:51 & 15.1 &  71.8 &  HD10380 & 07:45:02 &  1.2 &  1.9 &  1& 1 &\#3\\
08:13:45 & 15.1 &  71.7 &  HD10380 & 07:45:02 &  1.3 &  2.1 &  1& 0 &\#3\\
08:52:57 & 14.0 &  70.2 &  HD18322 & 09:18:41 &  1.4 &  1.7 &  1& 1 &\#3\\
08:56:48 & 13.9 &  70.0 &  HD18322 & 09:18:41 &  1.4 &  1.7 &  1& 0 &\#3\\
09:37:35 & 12.3 &  67.4 &  HD18322 & 09:18:41 &  1.7 &  1.4 &  0& - & -\\
\hline
\multicolumn{6}{l}{2007-10-08: G0H0} \T \B\\
05:09:01 & 29.4 &  71.2 &  HD10380 & 04:34:38 &  1.1 &  1.3 &  1 & 0 & \#7\\
05:13:07 & 29.6 &  71.3 &  HD10380 & 05:48:11 &  1.1 &  1.2 &  1& 1 & \#7\\
05:17:10 & 29.8 &  71.5 &  HD10380 & 05:48:11 &  1.1 &  1.1 &  1& 1 &\#7\\
05:21:25 & 30.0 &  71.6 &  HD10380 & 05:48:11 &  1.1 &  1.1 &  1& 0 &\#7\\
06:12:14 & 31.7 &  72.6 &  HD10380 & 05:48:11 &  1.1 &  1.5 &  1& 1 & \#8\\
06:15:59 & 31.7 &  72.7 &  HD10380 & 05:48:11 &  1.1 &  1.6 &  1& 0 & \#8\\
06:20:58 & 31.8 &  72.7 &  HD10380 & 06:45:43 &  1.1 &  1.7 &  1& 1 & \#8\\
06:31:48 & 31.9 &  72.8 &  HD10380 & 06:45:43 &  1.1 &  1.8 &  1& 0 & \#8\\
07:03:08 & 31.9 &  72.8 &  HD10380 & 06:45:43 &  1.1 &  1.7 &  0 &- &-\\
07:07:36 & 31.9 &  72.7 &  HD10380 & 06:45:43 &  1.1 &  1.7 &  1& 1 & \#8\\
07:11:30 & 31.8 &  72.7 &  HD10380 & 06:45:43 &  1.1 &  1.7 &  1& 0 & \#8\\
07:15:16 & 31.8 &  72.7 &  HD10380 & 07:44:12 &  1.1 &  1.6 &  1& 1 & \#8\\
07:19:22 & 31.7 &  72.6 &  HD10380 & 07:44:12 &  1.1 &  1.4 &  1& 0 &\#8\\
08:01:29 & 30.5 &  71.9 &  HD10380 & 07:44:12 &  1.2 &  1.0 &  1& 1 & \#7\\
08:05:26 & 30.3 &  71.8 &  HD10380 & 07:44:12 &  1.2 &  0.9 &  1& 0 & \#7\\
08:09:10 & 30.1 &  71.7 &  HD10380 & 08:28:09 &  1.3 &  1.0 &  0 &- &-\\
08:12:56 & 30.0 &  71.6 &  HD10380 & 08:28:09 &  1.3 &  1.0 &  1& 0 &\#7\\
08:16:46 & 29.8 &  71.5 &  HD10380 & 08:28:09 &  1.3 &  1.0 &  0 &- &-\\
08:57:45 & 27.4 &  69.8 &  HD18322 & 09:24:25 &  1.5 &  0.8 &  1& 1 &\#7\\
09:01:38 & 27.1 &  69.6 &  HD18322 & 09:24:25 &  1.5 &  0.9 &  1& 1 &\#7\\
09:05:35 & 26.8 &  69.4 &  HD18322 & 09:24:25 &  1.5 &  0.8 &  1& 0 &\#7\\
\hline
\multicolumn{6}{l}{2012-09-19: I1K0} \T \B\\
07:43:13 & 42.9 &  15.0 &  HD10380 & 07:29:15 &  1.1 &  0.9 &  1& 0 &\#10\\
08:01:30 & 43.3 &  16.8 &  HD18322 & 07:52:50 &  1.1 &  1.0 &  1& 0 &\#10\\
08:05:16 & 43.4 &  17.2 &  HD18322 & 08:16:55 &  1.1 &  1.1 &  1& 1 &\#10\\
08:09:03 & 43.5 &  17.5 &  HD18322 & 08:16:55 &  1.1 &  1.0 &  1& 1 &\#10\\
08:25:04 & 43.8 &  18.9 &  HD18322 & 08:16:55 &  1.1 &  0.8 &  1& 0 &\#10\\
08:29:02 & 43.9 &  19.3 &  HD18322 & 08:16:55 &  1.1 &  0.7 &  0 &- &-\\
08:36:16 & 44.1 &  19.9 &  HD18322 & 08:43:39 &  1.2 &  0.7 &  1& 1 &\#11\\
08:50:25 & 44.4 &  21.0 &  HD18322 & 08:43:39 &  1.2 &  0.7 &  1& 1 &\#11\\
08:52:46 & 44.5 &  21.2 &  HD18322 & 08:43:39 &  1.2 &  0.7 &  1& 1 &\#11\\
08:55:14 & 44.5 &  21.3 &  HD18322 & 08:43:39 &  1.2 &  0.6 &  1& 0 &\#11\\
08:57:31 & 44.6 &  21.5 &  HD18322 & 09:07:06 &  1.2 &  0.6 &  1& 1 &\#11\\
08:59:48 & 44.6 &  21.7 &  HD18322 & 09:07:06 &  1.2 &  0.6 &  1& 0 &\#11\\
\hline
\multicolumn{6}{l}{2012-09-20: G1I1} \T \B\\
06:01:26 & 39.8 &  35.3 &  HD10380 & 05:51:28 &  1.2 &  0.8 &  1& 0 &\#9\\
06:04:53 & 40.0 &  35.8 &  HD10380 & 06:16:38 &  1.2 &  0.8 &  1& 1 &\#9\\
06:08:31 & 40.2 &  36.2 &  HD10380 & 06:16:38 &  1.2 &  0.9 &  1& 1 &\#9\\
06:25:07 & 41.2 &  38.0 &  HD10380 & 06:16:38 &  1.1 &  0.9 &  1& 0 &\#9\\
06:28:35 & 41.4 &  38.4 &  HD10380 & 06:39:43 &  1.1 &  1.0 &  1& 1 &\#9\\
06:32:00 & 41.6 &  38.7 &  HD10380 & 06:39:43 &  1.1 &  1.0 &  1& 0 &\#9\\
\hline
\end{tabular}
\end{table*}

\begin{table*}
\caption{Log of observations: NGC1068}
\centering
\begin{tabular}{c c c c c c c c c c }
\hline\hline
Time & BL [m] & PA [deg] & Calibrator & Caltime &  Airmass & Seeing & Good? & Stacked  & Group\\
\hline
\hline
\multicolumn{6}{l}{2012-09-24: C1D0} \\
04:28:29 & 18.6 &   7.0 &  HD10380 & 04:20:28 &  1.4 &  0.5 &  1 & 0 &\#4\\
04:32:01 & 18.6 &   7.6 &  HD10380 & 04:39:44 &  1.4 &  0.5 &  1 & 1 &\#4\\
04:51:21 & 18.8 &  10.8 &  HD10380 & 04:39:44 &  1.3 &  0.6 &  1 & 0 &\#4\\
04:54:55 & 18.9 &  11.4 &  HD10380 & 05:03:59 &  1.3 &  0.6 &  1 & 0 &\#4\\
\hline
\multicolumn{6}{l}{2012-09-24: B2D0} \\
05:37:00 & 29.1 &  17.8 &  HD10380 & 05:29:55 &  1.2 &  0.6 &  1& 0 &\#6\\
05:40:32 & 29.2 &  18.3 &  HD10380 & 05:48:33 &  1.2 &  0.6 &  1& 1 &\#6\\
05:55:07 & 29.6 &  20.3 &  HD10380 & 05:48:33 &  1.2 &  0.6 &  1& 0 &\#6\\
05:58:42 & 29.7 &  20.8 &  HD10380 & 06:06:23 &  1.2 &  0.5 &  1& 0 &\#6\\
\hline
\multicolumn{6}{l}{2012-09-24: B2C1} \\
06:28:04 & 10.1 &  24.3 &  HD10380 & 06:20:35 &  1.1 &  0.6 &  1& 0 &\#1\\
06:32:04 & 10.2 &  24.8 &  HD10380 & 06:39:50 &  1.1 &  0.6 &  1& 0 &\#1\\
06:53:17 & 10.4 &  27.0 &  HD10380 & 07:04:27 &  1.1 &  0.6 &  1& 1 &\#1\\
06:56:48 & 10.4 &  27.3 &  HD10380 & 07:04:27 &  1.1 &  0.6 &  1& 0 &\#1\\
\hline
\multicolumn{6}{l}{2012-09-25: A1B2} \\
06:35:55 & 11.3 & 113.5 &  HD10380 & 06:47:45 &  1.1 &  0.8 &  1& 1 &\#2\\
06:39:43 & 11.3 & 113.5 &  HD10380 & 06:47:45 &  1.1 &  0.7 &  1& 0 &\#2\\
06:59:49 & 11.1 & 113.8 &  HD10380 & 07:10:26 &  1.1 &  0.7 &  1& 1  &\#2\\
07:03:20 & 11.1 & 113.9 &  HD10380 & 07:10:26 &  1.1 &  0.7 &  1& 0  &\#2\\
\hline
\multicolumn{6}{l}{2012-09-25: B2C1} \\
07:34:39 & 10.7 &  30.8 &  HD18322 & 07:28:07 &  1.1 &  0.6 &  1& 0 &\#1\\
07:38:25 & 10.8 &  31.0 &  HD18322 & 07:46:10 &  1.1 &  0.7 &  1& 1 &\#1\\
07:51:59 & 10.9 &  31.9 &  HD18322 & 07:46:10 &  1.1 &  0.9 &  1& 0 &\#1\\
07:55:31 & 10.9 &  32.1 &  HD18322 & 08:03:51 &  1.1 &  0.9 &  1& 0 &\#1\\
\hline
\multicolumn{6}{l}{2012-09-25: C1D0} \\
08:21:07 & 22.2 &  33.5 &  HD18322 & 08:15:06 &  1.2 &  0.6 &  1 & 0  &\#5\\
08:25:38 & 22.2 &  33.6 &  HD18322 & 08:35:25 &  1.2 &  0.6 &  0& - &- \\
08:27:22 & 22.2 &  33.7 &  HD18322 & 08:35:25 &  1.2 &  0.7 &  1& 1 & \#5\\
08:41:52 & 22.4 &  34.2 &  HD18322 & 08:35:25 &  1.2 &  0.7 &  1& 1 & \#5\\
08:45:34 & 22.4 &  34.4 &  HD18322 & 08:35:25 &  1.2 &  0.7 &  1& 0 & \#5\\
09:05:55 & 22.5 &  34.9 &  HD18322 & 09:00:33 &  1.3 &  0.7 &  1& 0  &\#5\\
09:10:13 & 22.5 &  35.0 &  HD18322 & 09:17:09 &  1.3 &  0.7 &  1& 1  &\#5\\
09:27:40 & 22.6 &  35.2 &  HD18322 & 09:17:09 &  1.4 &  0.7 &  1& 0  &\#5\\
09:30:02 & 22.6 &  35.2 &  HD18322 & 09:40:12 &  1.4 &  0.6 &  1& 1  &\#5\\
09:32:10 & 22.6 &  35.2 &  HD18322 & 09:40:12 &  1.4 &  0.7 &  0& - &- \\
09:34:20 & 22.6 &  35.2 &  HD18322 & 09:40:12 &  1.4 &  0.7 &  1& 0  &\#5\\
\hline
\end{tabular}
\end{table*}

\end{appendix}

\end{document}